\documentclass[12pt,A4paper]{article}
\usepackage{amsmath,amssymb,array,calc,rotating,epsfig,psfrag, amscd,caption2}
\usepackage[left=2cm,top=2cm,right=2cm,nohead]{geometry}
\newcommand{\beq}{\begin{equation}\begin{aligned}}
\newcommand{\eeq}{\end{aligned}\end{equation}}

\newcommand{\nn}{\nonumber}

\newcommand{\fc}{\phi^c}
\newcommand{\fd}{\phi^d}

\newcommand{\fo}{\phi_0}

\newcommand{\fua}{\phi_1^a}
\newcommand{\fub}{\phi_1^b}

\newcommand{\tix}{\tilde X}

\newcommand{\nc}{\newcommand}

\nc{\ra}{\rightarrow}
\nc{\lra}{\leftrightarrow}
\nc{\Ra}{\Rightarrow}
\nc{\LRa}{\Leftightarrow}
\nc{\blp}{{\big (}}
\nc{\brp}{{\big )}}
\nc{\Blp}{{\Big (}}
\nc{\Brp}{{\Big )}}
\nc{\bglp}{{\bigg (}}
\nc{\bgrp}{{\bigg )}}
\nc{\Bglp}{{\Bigg (}}
\nc{\Bgrp}{{\Bigg )}}
\nc{\slb}{{\rm [}}
\nc{\srb}{{\rm ]}}
\def\al{\alpha}

\nc{\veps}{\varepsilon}
\def\gam{\gamma}

\def\lam{\lambda}
\def\om{\omega}
\nc{\vphi}{\varphi}
\def\tha{\theta}
\def\sig{\sigma}

\def\Gam{\Gamma}

\def\Om{\Omega}
\def\Sig{\Sigma}

\nc{\bea}{\begin{eqnarray}}
\nc{\eea}{\end{eqnarray}}
\nc{\be}{\begin{equation}}
\nc{\ee}{\end{equation}}
\nc{\cA}{{\cal A}}
\nc{\cB}{ \cal B}

\nc{\cF}{{\cal F}}
\nc{\cG}{{\cal G}}

\nc{\cL}{{\cal L}}
\nc{\cM}{{\cal M}}
\def\N{{\cal N}}

\def\cO{{\cal O}}

\nc{\cQ}{{\cal Q}}
\nc{\cR}{{\cal R}}

\def\cV{{\cal V}}
\def\cV{{\cal V}}

\def\cZ{{\cal Z}}
\nc{\cQd}{\cQ^{\dagger}}
\nc{\cRd}{\cR^{\dagger}}
\nc{\BB}{{\mathbb B}}
\nc{\CC}{{\mathbb C}}
\nc{\DD}{{\mathbb D}}
\nc{\EE}{{\mathbb E}}
\nc{\FF}{{\mathbb F}}
\nc{\GG}{{\mathbb G}}
\nc{\HH}{{\mathbb H}}
\nc{\JJ}{{\mathbb J}}
\nc{\RR}{{\mathbb R}}
\nc{\PP}{{\mathbb P}}
\nc{\QQ}{{\mathbb Q}}
\nc{\ZZ}{{\mathbb Z}}
\nc{\calone}{{\mathbb 1}}
\nc{\half}{\frac{1}{2}}
\nc{\qrt}{\frac{1}{4}}
\nc{\del}{\partial}

\nc{\delbar}{\bar\partial}
\nc{\Spin}{\operatorname{Spin}}
\nc{\SO}{\operatorname{SO}}

\nc{\Sp}{{\rm Sp}}
\nc{\com}[2]{{ \left[ #1, #2 \right] }}
\nc{\acom}[2]{{ \left\{ #1, #2 \right\} }}
\nc{\rr}{\rightarrow}
\nc{\p}{\partial}
\nc{\LT}{{\LL_\T}}
\nc{\Tr}{{\rm Tr}}
\nc{\tr}{{\rm tr}}
\def\com#1#2{{ \left[ #1, #2 \right] }}
\def\acom#1#2{{ \left\{ #1, #2 \right\} }}
\nc{\Adag}{A^{\dagger}}
\nc{\AdagI}{A^{\dagger I}}
\nc{\AdagJ}{A^{\dagger J}}
\nc{\AdagK}{A^{\dagger K}}
\nc{\AdagL}{A^{\dagger L}}
\nc{\AdagM}{A^{\dagger M}}
\nc{\Bdag}{B^{\dagger}}
\nc{\BdagI}{B^{\dagger}_I}
\nc{\BdagJ}{B^{\dagger}_J}
\nc{\BdagK}{B^{\dagger}_K}
\nc{\BdagL}{B^{\dagger}_L}
\nc{\BdagM}{B^{\dagger}_M}
\nc{\Cdag}{C^{\dagger}}
\nc{\CdagI}{C^{\dagger I}}
\nc{\CdagJ}{C^{\dagger J}}
\nc{\CdagK}{C^{\dagger K}}
\nc{\Ddag}{D^{\dagger}}
\nc{\DdagI}{D^{\dagger I}}
\nc{\DdagJ}{D^{\dagger J}}
\nc{\DdagK}{D^{\dagger K}}
\nc{\ttha}{\tilde{\theta}}
\nc{\tphi}{\tilde{\phi}}
\nc{\tsig}{\tilde{\sig}}
\nc{\tom}{\tilde{\om}}
\nc{\tlam}{\tilde{\lam}}
\nc{\tSig}{\widetilde{\Sig}}
\nc{\tPhi}{\tilde{\Phi}}
\nc{\tPhibar}{\ol{\tPhi}}
\nc{\tPi}{\tilde{\Pi}}
\nc{\tpsi}{\tilde{\psi}}
\nc{\tPsi}{\tilde{\Psi}}
\nc{\tgam}{\tilde{\gam}}
\nc{\tGam}{\tilde{\Gam}}
\nc{\txi}{\tilde{\xi}}
\nc{\tXi}{\tilde{\Xi}}
\nc{\tb}{\tilde b}
\nc{\tc}{\tilde c}
\nc{\te}{\tilde e}
\nc{\tf}{\tilde f}
\nc{\tg}{\tilde g}
\nc{\tj}{\tilde j}
\nc{\tp}{\widetilde{p}}
\nc{\tq}{\widetilde{q}}
\nc{\ts}{{\tilde s}}
\nc{\tu}{{\tilde u}}
\nc{\tv}{{\tilde v}}
\nc{\tw}{{\tilde w}}
\nc{\tx}{{\tilde x}}
\nc{\ty}{{\tilde y}}
\nc{\tz}{\tilde z}
\nc{\tA}{{\tilde A}}
\nc{\tAbar}{{\ol \tA}}
\nc{\tD}{{\tilde D}}
\nc{\tE}{{\tilde E}}
\nc{\tG}{{\tilde G}}
\nc{\tH}{{\tilde H}}
\nc{\tJ}{{\tilde J}}
\nc{\tJbar}{{\ol {\tilde J}}}
\nc{\tM}{{\tilde M}}
\nc{\tN}{{\tilde N}}
\nc{\tP}{{\tilde P}}
\nc{\tQ}{{\tilde Q}}
\nc{\tR}{{\tilde R}}
\nc{\tS}{\tilde{S}}
\nc{\tF}{\tilde{{\cal F}}}
\nc{\tX}{\widetilde{X}}
\nc{\tcZ}{\tilde{\cZ}}
\nc{\tcZbar}{\ol{\tcZ}}
\nc{\hb}{\hat b}
\nc{\hc}{\hat c}
\nc{\hd}{\hat d}
\nc{\he}{\hat e}
\nc{\hf}{\hat f}
\nc{\hg}{\hat g}
\nc{\hh}{\hat h}
\nc{\hp}{\hat p}
\nc{\hv}{\hat v}
\nc{\hw}{\hat w}
\nc{\hx}{\hat x}
\nc{\hy}{\hat y}
\nc{\hz}{\hat z}
\nc{\hA}{\widehat{A}}
\nc{\hE}{\widehat{E}}
\nc{\hF}{\widehat{F}}
\nc{\hH}{\widehat{H}}
\nc{\hJ}{\widehat{J}}
\nc{\tK}{\widetilde{K}}
\nc{\hM}{\widehat M}
\nc{\hV}{\widehat V}
\nc{\hcV}{\widehat \cV}

\nc{\ha}{\widehat \alpha}
\nc{\hphi}{\hat{\phi}}
\nc{\hpsi}{\hat{\psi}}
\nc{\hgam}{\hat{\gam}}
\nc{\hPhi}{\hat{\Phi}}
\nc{\hPsi}{\hat{\Psi}}
\nc{\hGam}{\hat{\Gam}}

\nc{\w}{\wedge}

\nc{\vb}{\vec b}
\nc{\vc}{\vec c}
\nc{\vd}{\vec d}
\nc{\ve}{\vec e}
\nc{\vf}{\vec f}
\nc{\vg}{\vec g}
\nc{\vh}{\vec h}
\nc{\vp}{\vec p}
\nc{\vq}{\vec q}
\nc{\vr}{\vec r}
\nc{\vs}{\vec s}
\nc{\vv}{\vec v}
\nc{\vw}{\vec w}
\nc{\vx}{\vec x}
\nc{\vy}{\vec y}
\nc{\vz}{\vec z}

\nc{\ol}{\overline}
\nc{\abar}{\ol{a}}
\nc{\bbar}{\ol{b}}
\nc{\cbar}{\ol{c}}
\nc{\dbar}{\ol{d}}
\nc{\ebar}{\ol{e}}
\nc{\ibar}{\ol{\imath}}
\nc{\jbar}{\ol{\jmath}}
\nc{\kbar}{\ol{k}}
\nc{\lbar}{\ol{l}}
\nc{\mbar}{\ol{m}}
\nc{\nbar}{\ol{n}}
\nc{\pbar}{\ol{p}}
\nc{\qbar}{\ol{q}}
\nc{\ubar}{\ol{u}}
\nc{\vbar}{\ol{v}}
\nc{\wbar}{\ol{w}}
\nc{\xbar}{\ol{x}}
\nc{\ybar}{\ol{y}}
\nc{\zbar}{\ol{z}}

\nc{\Abar}{\ol{A}}
\nc{\Bbar}{\ol{B}}
\nc{\Cbar}{\ol{C}}
\nc{\Dbar}{\ol{D}}
\nc{\Ebar}{\ol{E}}
\nc{\Jbar}{\ol{J}}
\nc{\Kbar}{\ol{K}}
\nc{\Lbar}{\ol{L}}
\nc{\Nbar}{\ol{N}}
\nc{\Pbar}{\ol{P}}
\nc{\Qbar}{\ol{Q}}
\nc{\Rbar}{\ol{R}}
\nc{\Sbar}{\ol{S}}
\nc{\Tbar}{\ol{T}}
\nc{\Ubar}{\ol{U}}
\nc{\Vbar}{\ol{V}}
\nc{\Wbar}{\ol{W}}
\nc{\Xbar}{{\overline X}}
\nc{\Ybar}{{\overline Y}}
\nc{\Zbar}{{\overline Z}}
\nc{\cZbar}{{\overline \cZ}}

\nc{\epsbar}{\ol{\epsilon}}
\nc{\lambar}{\ol{\lambda}}
\nc{\psibar}{\ol{\psi}}
\nc{\Psibar}{\ol{\Psi}}
\nc{\phibar}{\ol{\phi}}
\nc{\Phibar}{\ol{\Phi}}
\nc{\chibar}{\ol{\chi}}
\nc{\mubar}{\ol{\mu}}
\nc{\nubar}{\ol{\nu}}
\nc{\rhobar}{\ol{\rho}}
\nc{\ombar}{\ol{\om}}
\nc{\Ombar}{\ol{\Om}}

\nc{\Dthb}{\ol{\rm D3}}

\nc{\gdot}{\dot{g}}
\nc{\xdot}{\dot{x}}
\nc{\ydot}{\dot{y}}
\nc{\sinp}{s_{\phi}}
\nc{\cosp}{c_{\phi}}
\nc{\tanp}{t_{\phi}}
\nc{\spone}{s_{\phi_1}}
\nc{\cpone}{c_{\phi_1}}
\nc{\tpone}{t_{\phi_1}}
\nc{\sptwo}{s_{\phi_2}}
\nc{\cptwo}{c_{\phi_2}}
\nc{\tptwo}{t_{\phi_2}}
\nc{\spth}{s_{\phi_3}}
\nc{\cpth}{c_{\phi_3}}
\nc{\tpth}{t_{\phi_3}}

\nc{\csch}{{\rm csch}}

\nc{\bah}{{\mathbf {\hat{A}}}}
\nc{\bX}{{\mathbf X}}
\nc{\ba}{{\bf a}}
\nc{\bb}{{\bf b}}
\nc{\bc}{{\bf c}}
\nc{\bd}{{\bf d}}
\nc{\bg}{{\bf g}}
\nc{\bk}{{\bf k}}
\nc{\bl}{{\bf l}}
\nc{\bm}{{\bf m}}
\nc{\bn}{{\bf n}}
\nc{\bo}{{\bf o}}
\nc{\bp}{{\bf p}}
\nc{\bq}{{\bf q}}
\nc{\br}{{\bf r}}
\nc{\bs}{{\bf s}}
\nc{\bt}{{\bf t}}
\nc{\bu}{{\bf u}}
\nc{\bv}{{\bf v}}
\nc{\bw}{{\bf w}}
\nc{\bx}{{\bf x}}
\nc{\by}{{\bf y}}
\nc{\bz}{{\bf z}}
\nc{\bom}{{\bf \om}}
\nc{\bombar}{{\mathbf \ombar}}
\nc{\bPhi}{{\bf \Phi}}

\nc{\rma}{{\rm a}}
\nc{\rmb}{{\rm b}}
\nc{\rmc}{{\rm c}}
\nc{\rmd}{{\rm d}}
\nc{\rmg}{{\rm g}}
\nc{\rk}{{\rm k}}
\nc{\rml}{{\rm l}}
\nc{\rmm}{{\rm m}}
\nc{\rmn}{{\rm n}}
\nc{\rmo}{{\rm o}}
\nc{\rmp}{{\rm p}}
\nc{\rmq}{{\rm q}}
\nc{\rmr}{{\rm r}}
\nc{\rms}{{\rm s}}
\nc{\rmt}{{\rm t}}
\nc{\rmu}{{\rm u}}
\nc{\rmv}{{\rm v}}
\nc{\rmw}{{\rm w}}
\nc{\rmx}{{\rm x}}
\nc{\rmy}{{\rm y}}
\nc{\rmz}{{\rm z}}

\nc{\vol}{{\rm vol}}
\nc{\dal}{\dot{\al}}
\nc{\thadot}{\dot{\tha}}
\nc{\thab}{\bar{\theta}}
\nc{\thal}{\theta^{\al}}
\nc{\thdal}{\bar{\theta}^{\dal}}

\nc{\thsigthm}{\tha \sigma^m \thab}
\nc{\thsigthn}{\tha \sigma^n \thab}

\nc{\Dal}{D_{\al}}
\nc{\Ddal}{\bar{D}_{\dal}}
\nc{\CDal}{{\cal D}_{\al}}
\nc{\CDdal}{\bar{\cal D}_{\dal}}

\nc{\eq}[1]{(\ref{#1})}
\nc{\non}{\nonumber}
\begin{document}
%%%%%%%%%%%%%%%%%%%%%%%%%%%%%%%%%%%%
%%%%%%%%%%%%%%%%%%%%%%%%%%%%%%%%%%%%
\rightline{IPhT-T09/237}

\begin{center}
\vskip 2 cm
{\Large \bf On the Existence of Meta-stable Vacua\\
\vskip 3mm
in Klebanov-Strassler}\\

\vskip 1.25 cm
{Iosif Bena, Mariana Gra\~na  and Nick Halmagyi}\\

\vskip 5mm
Institut de Physique Th\' eorique, \\
CEA Saclay, CNRS-URA 2306, \\
91191 Gif sur Yvette, France \\
\vskip 5mm

iosif.bena, \  mariana.grana, \ nicholas.halmagyi@cea.fr.

\end{center}

\begin{abstract}
We solve for the complete space of linearized deformations of the
  Klebanov-Strassler background consistent with the symmetries
  preserved by a stack of anti-D3 branes smeared on the $S^3$ of the
  deformed conifold.  We find that the only solution whose UV physics
  is consistent with that of a perturbation produced by anti-D3 branes
  must have a singularity in the infrared, coming from NS and RR
  three-form field strengths whose energy density diverges. If this
  singularity is admissible, our solution describes the backreaction
  of the anti-D3 branes, and is thus likely to be dual to the
  conjectured metastable vacuum in the Klebanov-Strassler field
  theory. If this singularity is not admissible, then our analysis
  strongly suggests that anti-D3 branes do not give rise to metastable
  Klebanov-Strassler vacua, which would have dramatic consequences for
  some string theory constructions of de Sitter space. Key to this
  result is a simple, universal form for the force on a probe D3-brane
  in our ansatz.
  
\end{abstract}
\newpage
\tableofcontents

\section{Introduction}

Over the past few years the analysis of metastable vacua of
supersymmetric field theories has begun to play an important role in
phenomenological model building, as these vacua can bypass many of
the
familiar problems associated with old-fashioned dynamical
supersymmetry breaking (DSB). Such meta-stable vacua were first found
for massive ${\cal N} = 1$ $SU(Nc)$ SQCD in the free-magnetic phase
\cite{Intriligator:2006dd}, and soon thereafter in quite a large
number of similar theories.

An obvious question posed by the ubiquitousness of metastable vacua in
SQCD theories is whether such vacua also exist in the string-theory
configurations whose low-energy physics yields these theories. These
constructions involve D-branes and NS5 branes in type IIA string
theory, and are generically referred to as MQCD \cite{Witten:1997ep}.
Another obvious question is whether such metastable vacua exist in the
type IIB supergravity backgrounds such as \cite{Klebanov:2000hb} that
are dual to some of the SQCD theories via the gauge-gravity ($AdS$-CFT)
correspondence.

For MQCD theories one can construct putative non-supersymmetric brane
configurations whose low-energy physics matches that of the SQCD
metastable vacuum. At $g_s=0$ these brane configurations have the same
asymptotics as the configuration describing the MQCD supersymmetric
vacua. However, for $g_s \neq 0$ these non-supersymmetric
configurations do not have the same asymptotics as the MQCD
supersymmetric branes, and hence do not describe a metastable vacuum
of a supersymmetric theory (where supersymmetry is broken
dynamically), but rather a non-supersymmetric vacuum of a
nonsupersymmetric theory \cite{Bena:2006rg}. The fact that the
putative MQCD non-supersymmetric and supersymmetric configurations are
not vacua of the same theory can also be ascertained by computing the
tunneling probability between them: for $g_s \neq 0$ this probability
is zero because the branes differ by an infinite amount, while for
$g_s = 0$ this probability is zero because the branes have infinite
mass.

The key reason for which the susy and non-susy brane configurations do
not have the same asymptotics comes from the phenomenon of brane
bending: essentially all MQCD brane descriptions of four-dimensional
theories contain D4 branes ending on NS5 branes; since the end of D4
brane is codimension 2 inside the NS5 brane, it sources an NS5
worldvolume field that is logarithmic at infinity, and that causes the
NS5 brane to bend logarithmically. For the MQCD dual of $\N=2$ SQCD,
this bending encodes the logarithmic running of the coupling constant
with the energy, characteristic of four-dimensional gauge theories
\cite{Witten:1997ep}. The D4 branes in the deep infrared of the susy
and non-susy MQCD brane configurations end on the NS5 branes in
different fashions, and source fields of different kinds. Since these
fields do not decay at infinity, but grow logarithmically, the tiny
infrared difference is transposed into a log-growing difference in the
UV, which is responsible for the susy an non-susy configurations not
being vacua of the same theory.

It is also interesting to note that for three-dimensional gauge
theories engineered using D3 branes and NS5 branes in type IIB string
theory, this phenomenon does not happen. The D3 branes end in
codimension-three sources on the NS5 branes, which source fields that
decay like $1/r$ at infinity. This bending encodes the running of
coupling constants in three dimensions, which is linear in the inverse
of the energy. Since the difference between the fields of the
non-supersymmetric and supersymmetric brane configurations decays at
infinity, the two do correspond to vacua of the same theory, and the
non-supersymmetric configuration is really a 3D MQCD metastable
vacuum.

Thus, the mechanism that ensures the absence of metastable vacua in
four-dimensional MQCD appears to rely on just one ingredient: the
existence of logarithmic modes that are sourced differently in the
infrared for the BPS and non-BPS configurations. As these modes are
present in four-dimensional but not in three dimensional gauge theories, it is
only four-dimensional MQCD which does not have metastable vacua,
three-dimensional MQCD does have such vacua.

In this paper we will analyze whether a similar mechanism is at work
in gravity duals of $\N=1$ four-dimensional gauge theories. As the
coupling constants of these theories run logarithmically with the
energy, all their gravity duals must contain fields that depend
logarithmically on the $AdS$ radius, at least
asymptotically\footnote{In the well-known Klebanov-Strassler solution
  \cite{Klebanov:2000hb} such a field is given by the integral of the
  NS B-field on the 2-cycle of the conifold.}.  It is possible that a
putative metastable configuration in the infrared couples to these
logarithmic modes in different manner than the supersymmetric
configuration; hence the non-BPS solution would differ in the
ultraviolet from the BPS solution by one or more non-normalizable
modes. If this indeed happens, then the non-BPS solution will not be
dual to a metastable DSB vacuum of the supersymmetric field theory.
Instead, it will describe a non-supersymmetric vacuum of a
non-supersymmetric theory obtained by perturbing the Lagrangian of the
supersymmetric theory. As a consequence, supersymmetry breaking will
not be dynamical but explicit.

The best-known candidate for the IIB gravity dual of a metastable
gauge theory vacuum has been proposed by Kachru, Pearson and Verlinde
(KPV) \cite{Kachru:2002gs}, who argued that a collection of $P$
anti-D3 branes placed at the bottom of the Klebanov-Strassler (KS)
solution with $M$ units of 3-form flux can decay into a BPS solution
corresponding to the KS solution with $M-P$ D3 branes at the bottom of
the throat. In particular, if $P=M$ this collection of anti-D3 branes
could decay into the smooth, source-free KS solution. The argument
presented by \cite{Kachru:2002gs} was in the probe approximation and
it important to extend this to a backreacted supergravity solution.
Finding the backreacted solution would allow one to establish whether
this configuration is indeed a metastable vacuum of a supersymmetric
theory, or it is rather a non-supersymmetric vacuum of a
non-supersymmetric theory. This necessity is particularly evident,
given that in MQCD there exist non-backreacted brane configurations
that are candidates for meta-stable vacua
\cite{Ooguri:2006bg,Franco:2006ht,Bena:2006rg} but in fact break
supersymmetry explicitly \cite{Bena:2006rg} once backreaction is taken
into account\footnote{Worldsheet evidence in support of this MQCD
  analysis was given in \cite{Murthy:2007qm} and furthermore, both the
  MQCD dual of the configuration of \cite{Kachru:2002gs} (analyzed in
  \cite{Mukhi:2000dn}) as well as other duals of KPV-like
  configurations \cite{Argurio:2006ny}, suffer from the same problems
  in the UV, at least in certain regime of parameters. }.

Note that this question is in no way settled by the results currently
in the literature. For example the ultraviolet perturbation expansion
around the Klebanov-Tseytlin \cite{Klebanov:2000nc} solution presented
in \cite{DeWolfe:2008zy} assumes {\it a-priori} that the anti-D3
branes source only a normalizable mode, and uses the consistency of
the result (in particular they use the fact that the force on a probe
D3 brane is not inconsistent with that calculated in
\cite{Kachru:2003sx}) to argue that such an assumption is correct.
However as we will show in Section \ref{forcesection}, all
$SU(2)\times SU(2)\times \ZZ_2$-invariant first-order perturbations
around the Klebanov-Strassler solution, whether normalizable or
non-normalizable, give a force on a probe D3 brane that is either zero
or has the precise behavior predicted in \cite{Kachru:2003sx}. Hence
this force calculation alone will not determine the normalizability or
non-normalizabilty of the modes sourced by an anti-D3 brane. In
addition, we will point out that subleading perturbations of the
(singular) Klebanov-Tseytlin geometry do not in fact correspond to
perturbations of the (regular) Klebanov-Strassler geometry and as such
the Klebanov-Tseytlin geometry is an inappropriate arena for studying
these issues. 

One can argue that because an anti-D3 brane is a point source in a
six-dimensional space, the fields sourced by such a brane should decay
at infinity as $1/r^4$, and thus could not correspond to
non-normalizable modes. As we will show in Section \ref{UVtoIR} by
explicit calculations, this intuition is not correct even under the
weakest of assumptions about the backreacted solution sourced by
anti-D3 branes, which must decay at least as strongly as $1/r^3$.
Furthermore, as we have argued
above, both the Klebanov-Strassler solution, as well as other
backgrounds dual to four-dimensional gauge theories whose coupling
constants run logarithmically with the energy, contain supergravity
modes that behave asymptotically as $\log r$. If the anti-D3 brane in
the deep infrared couples to such a mode, then the backreacted
solution would differ from the BPS solution by a non-normalizable
mode, and hence would not correspond to a metastable DSB vacuum.

It is also interesting to note that the only known fully-backreacted supergravity
solution corresponding to a DSB metastable vacuum, constructed by
Maldacena and N\u astase \cite{Maldacena:2001pb}, describes a vacuum of a
{\it 2+1 dimensional} gauge theory, for which none of the arguments
presented above applies. As we have discussed above, these theories
also have MQCD metastable DSB vacua, and their coupling constants are
linear in the inverse of the energy, and hence do not correspond to
bulk fields that grow in the UV and that could amplify the tiny
infrared difference between the BPS and non-BPS solutions into a
non-normalizable UV mode.

Our purpose in this paper is to establish the nature of the fields that
are sourced by anti-D3 branes placed at the bottom of the
Klebanov-Strassler solution, and to see whether these branes give rise
to normalizable or non-normalizable modes. To do this in generality
one would need to find the fully-backreacted solution corresponding to
anti-D3 branes in the warped deformed conifold. Since these branes are
point sources in the internal space, and since moreover they polarize
(by the Myers effect \cite{Myers:1999ps}) into D5 or NS5 branes
\cite{Kachru:2002gs}, the fields of such a non-supersymmetric solution
would depend on three or four variables. Finding such a solution is
clearly beyond hope with current technology.

One way to simplify the problem is to smear the anti-D3 branes on the
$S^3$ of the deformed conifold, so that the resulting backreacted
solution preserves the $SU(2)\times SU(2) \times \ZZ_2$ symmetry of
the original solution. The ansatz for the most general solution
preserving this isometry was written down by Papadopoulous and
Tseytlin \cite{Papadopoulos:2000gj}, who reduced the supergravity
equations of motion to second-order ordinary differential equations
for eight scalar functions of the ``radial'' variable $\tau$ of the KS
solution.

One might object that this smearing washes down important features of
the anti-D3 branes that give the putative metastable vacuum. Indeed,
as discussed before, when these anti-D3 branes are
together they acquire extra dipole moments by polarization, and hence
couple to more supergravity fields than the smeared D3 branes.
Furthermore, because of the broken supersymmetry one can argue that
the anti-D3 branes cannot remain smeared on the $S^3$, but will
eventually condense into a single-center configuration. This
condensation however takes place over a time that can be tuned to be
parametrically large, and hence does not affect the fact the that
smeared anti-D3 branes give rise to consistent supergravity solution.
Furthermore, as the smeared anti-D3 branes couple to {\it less} fields
than the localized polarized ones, their potential for sourcing a
potentially-dangerous mode that is non-normalizable in the ultraviolet
will be much lower than for the localized and polarized anti-D3
branes. Hence, if the smeared branes give rise to non-normalizable
modes, the localized branes will very likely do the same; however, if
the smeared branes do not source non-normalizable modes, this does not
exclude at all the possibility that the localized polarized branes
will source them.

Even after smearing the anti-D3 branes, solving the system of eight
coupled nonlinear differential equations is no easy task. To make
the general problem tractable one must use the fact that the fields
sourced by the anti-D3 branes are subdominant compared to the fields
of the background, at least in the region far away from the tip.
Hence, one can study these fields by performing a first-order
perturbation expansion around the supersymmetric KS solution.

Fortunately, as shown in detail in \cite{Borokhov:2002fm}, the
second-order equations satisfied by the perturbations of the eight
scalar functions of the PT ansatz around the KS solution, factorize
into sixteen first-order equations of which eight form a closed
system.  Note that this factorization does not throw away any of the
modes: both the eight second-order equations and the sixteen
first-order ones have sixteen integration constants. In fact, as we
will show in Section \ref{sec:zeroenergy}, one of these integration
constants is a gauge artifact and another one must be fixed to zero in
order for the solutions to the sixteen equations to give a solution to
Einstein's equations. Hence, the class of perturbative solutions that
we construct depends on fourteen physically-relevant integration
constants!

Before beginning it is worth discussing the implication of the
potential existence of a non-normalizable mode sourced by anti-D3
branes at the bottom of the Klebanov-Strassler solution. Such
anti-branes are part of the ``staple diet'' of string phenomenology
and string cosmology constructions: they are a crucial ingredient in
the KKLT construction of de Sitter space \cite{Kachru:2003aw} and in
the KKLMMT model for string theory inflation \cite{Kachru:2003sx}.
These constructions involve placing anti-D3 branes in a
Klebanov-Strassler throat that is glued to an ambient Calabi-Yau. The
anti-branes lift the energy of the $AdS$ vacuum giving rise to
de-Sitter vacua and can create a potential for a probe D3 brane that
drives inflation. Furthermore, one generally assumes that the extra
energy the branes bring is parametrically redshifted in the KS throat,
and can give rise to mass hierarchies.

A non-normalizable mode would ``climb'' up all the way to the end of
the Klebanov-Strassler throat, and will probably affect the gluing of
this throat to the ambient Calabi-Yau. However, as this gluing is not
understood, it is premature to say whether it will be negatively
affected by the non-normalizable mode. Nevertheless, if a
non-normalizable mode is present, its energy will not be localized at
the bottom of the KS throat, but rather up the throat, at the junction
between this throat and the ambient Calabi-Yau. Hence, the extra
energy brought about by an anti-D3 brane will contain both a
redshifted contribution, coming from the bottom of the KS throat, and
a non-redshifted contribution, from the junction region.

It would be interesting to explore whether this extra contribution to
the energy could invalidate the KKLT construction of de Sitter vacua
in string theory \cite{Kachru:2003aw}. We would like to remind the
reader that in this construction one adds anti-D3 branes to the bottom
of a KS throat in order to lift the $AdS$ vacuum to de Sitter. The
contribution of these anti-D3 branes to the total energy needs to be
small in order not to overrun the quantum corrections that are needed
to stabilize the K\"ahler moduli. In \cite{Kachru:2003aw} it was
assumed that the contribution of these anti-D3 branes can be made
exponentially small by placing them in an arbitrarily-long KS throat.
Nevertheless, if anti-D3 branes source non-normalizable modes, the
energy they bring cannot be made exponentially small by lengthening
the throat. Hence, in some circumstances and for certain
non-normalizable modes, this energy may be too large to allow the
K\"ahler moduli to be stabilized. This without doubt negatively
impacts the existence of large parts of the landscape of de Sitter
vacua in string theory.

On the other hand, as we will show in Section \ref{forcesection}, the
force felt by a probe D3 brane in the background sourced by anti-D3
branes does not depend at all on the presence of extra
non-normalizable modes. It is always
of the form $1/r^5$, exactly as needed in the KKLMMT model for
inflation in string theory \cite{Kachru:2003sx}.

%%%%%%%%%%%%%%%%%%%%%%%%%%%%%%%%%%
\subsection{Strategy and Results} \label{sec:strategy}

Our goal is to identify the modes sourced by anti-D3 branes, smeared
in an $SU(2)\times SU(2)\times \ZZ_2$-invariant way at the tip of the
warped deformed conifold. This involves solving the eight second-order
coupled nonlinear differential equations that govern the metric and
fluxes of the Papadopoulous-Tseytlin ansatz
\cite{Papadopoulos:2000gj}.  We simplify the problem by solving this
system of equations perturbatively, treating the anti-D3 branes as a
perturbation of the Klebanov-Strassler solution, and using the fact
that the second-order equations factorize into sixteen first-order
linear ones out of which form a closed system, that are relatively
simpler to solve \cite{Borokhov:2002fm}. One of the main achievements
of our paper is that we are able to solve these equations exactly in
terms of integrals.

Having set up this rather heavy computational machinery, we then
identify which parameters correspond to non-normalizable modes and
which do not. Our first strategy is to enforce the correct UV boundary
conditions appropriate for the addition of $\Nbar$ anti-D3 branes in
the KS background and investigate the resulting IR behavior.  This
amounts to requiring that there are $-\Nbar$ units of additional
Maxwell charge, that there is a non-zero force on a probe D3-brane in
the UV and the charge is the only non-normalizable mode which we allow
for.  We find that the resulting infrared geometry contains fields
that diverge in a subtle manner which seems to be incompatible with
anti-D3 brane sources. 

As we will discuss in detail in subsection (\ref{Singularities})
these singularities do not appear to have a distinct physical origin,
as opposed to say singularities that come from explicit D-brane
sources. On the other hand, they are milder than the latter, as their
divergent energy density integrates to a finite action. However, even
such mild singularities must normally be excluded in order for the
gauge/gravity duality to make sense\footnote{The best example of an
  unphysical singularity whose action is finite is the negative-mass
  Schwarzschild solution.}.  Whether this singularity is to be allowed
or not gives rise to two very distinct conclusions.  If the
singularity is to be allowed, then we argue that we have found the
correct non-supersymmetric linearized solution for anti-D3 branes in
KS. If the singularity is to be deemed inadmissible then we argue that
there is no solution which can interpolate between the KS boundary
conditions in the UV and anti-D3 brane boundary conditions in the IR.

Our second strategy is to enforce anti-D3 brane boundary conditions in the infrared
while disallowing the aforementioned singularities and examine the
consequences for the ultraviolet. We show that the ultraviolet of this
solution cannot match that of the KS solution with no non-normalizable
modes turned on. As a consequence we argue that in this case the
correct UV behavior is a large deformation of the KS boundary
conditions.

There can of course only be one correct solution for anti-D3 branes in
KS and since the two conclusions we have offered here differ so wildly
from each other that it is of utmost importance to rigorously settle
the issue of admissibility of the singularity which we have uncovered.
We present various arguments for and against this singularity in the
discussion section.

This paper is organized as follows: In section 2 we set up the
perturbation theory following \cite{Borokhov:2002fm}. In section 3 we
compute the force on a probe D3-brane in our supergravity ansatz and
show that is has a simple universal form. In section 4 we solve for
the modes and present their perturbative expansions. In section 5 we
discuss the boundary conditions for D3 and anti-D3 branes in our
backgrounds. In section 6 we attempt to construct the backreacted
solution corresponding to a stack of smeared anti-D3 branes in the KS
geometry. We conclude and discuss the implications of our results in
section \ref{disc}. Appendix \ref{conventions} presents our
conventions and appendix \ref{xiasympt} shows the asymptotic expansion
of the auxiliary fields.

%%%%%%%%%%%%%%%%%%%%%%%%%%%%%%%%%%
 \section{Perturbations around a supersymmetric solution}
 %%%%%%%%%%%%%%%%%%%%%%%%%%%%%%%%%%

 To solve the IIB supergravity equations perturbatively around a
 supersymmetric solution, we will use the method developed by Borokhov
 and Gubser \cite{Borokhov:2002fm}, who reduce the set of $n$ second
 order equations for $n$ fields $\phi^a$ depending on a single
 (radial) variable $\tau$ to $2n$ first order equations for $\phi^a$
 and their ``canonical conjugate variables" $\xi_a$. For this
 procedure to work, it is essential that the symmetries of the
 supergravity problem are sufficiently strong so as to just allow a
 dependance of the fields on a single, radial, variable.

 Since we are perturbing around a known solution, the $n$ linear
 second order equations from linearizing Einstein's equation can of
 course be rewritten as a system of $2n$ first order equations without
 losing any information\cite{See \cite{Berg:2006xy} for an interesting calculation using this technique.}. While this can obviously always be done, the
 formalism of \cite{Borokhov:2002fm} allows eight of these equations
 to form a closed subsystem when perturbing around a supersymmetric
 solution.  Related perturbation theory working directly with the
 second order equations appeared in \cite{Aharony:2005zr}.

%%%%%%%%%%%%%%%%%%%%%%%%%%%%%%%%%%
\subsection{The First Order Formalism}
The starting point is the one dimensional Lagrangian
\be
\cL=-\half G_{ab} \frac{d \phi^a}{ d \tau}  \frac{d \phi^b}{ d \tau}  - V(\phi) \label{Lag1}
\ee
 which we require to have the simplifying property that it can be written in terms of a superpotential
 \be
 \cL= -\half G_{ab} \Blp \frac{d \phi^a}{ d \tau}  -\half G^{ac} \frac{\del W}{ \del \phi^c} \Brp \Blp\frac{d \phi^a}{ d \tau}  -\half G^{ac} \frac{\del W}{ \del \phi^c}  \Brp - \half \frac{\del W}{\del \tau},
 \ee
 where
 \be
 V(\phi) = \frac{1}{8} G^{ab }\frac{\del W}{ \del \phi^a}\frac{\del W}{ \del \phi^b}.
 \ee

 The fields $\phi^a$ are expanded around their supersymmetric background value $\phi^a_0$ (which will correspond in our case to the Klebanov-Strassler solution \cite{Klebanov:2000hb})
 \beq
 \label{split}
 \phi^a=\phi^a_0+\phi^a_1(X) + {\cal O}(X^2)\ ,
 \eeq
where $X$ represents the set of perturbation parameters, and $\phi^a_1$ is linear in them.
The supersymmetric solution $\phi^a=\phi^a_0$ satisfies the gradient flow equations
\beq
\frac{d\phi^a}{d\tau}-\frac12 G^{ab} \frac{\partial W}{\partial \phi^b} =0\ ,\label{gradflow}
\eeq
while the deviation from the gradient flow equations for the perturbation $\phi_1^a$ is measured
by the conjugate functions $\xi_a$, given by
\beq
\label{xidef}
\xi_a\equiv G_{ab}(\phi_0) \left(\frac{d\phi_1^b}{d\tau} -M^b_{\ d} (\phi_0) \phi_1^d \right) \ , \qquad M^b{}_d\equiv\frac12 \frac{\partial}{\partial \fd} \left( G^{bc} \frac{\partial W}{\partial \fc} \right) \ .
\eeq
The $\xi_a$ are linear in the expansion parameters $X$, hence they are of the same order as the $\phi_1^a$, and when all the $\xi_a$ vanish the deformation is supersymmetric.

The main point of this construction is that the equations of motion reduce to a set of first order linear equations for $(\xi_a,\phi^a)$:
\bea
\frac{d\xi_a}{d\tau} + \xi_b M^b{}_a(\fo) &=& 0,  \label{xieq} \\
\frac{d\fua}{d\tau} - M^a{}_b(\fo) \fub &=& G^{ab} \xi_b \label{phieq} \ .
\eea
Note that equations (\ref{phieq}) are just a rephrasing of  the definition of the $\xi_a$ in \eq{xidef}, while Eqs. (\ref{xieq})
imply the equations of motion \cite{Borokhov:2002fm}.

%%%%%%%%%%%%%%%%%%%%%%%%%%%%%%%%%%
 \subsection{Papadopoulos-Tseytlin ansatz for the perturbation}
%%%%%%%%%%%%%%%%%%%%%%%%%%%%%%%%%%

 The KS background has $SU(2)\times SU(2)\times\ZZ_2$ symmetry. We are interested in a solution for the backreaction of a smeared set of anti-D3 branes and we have the liberty to smear these branes without breaking the $\ZZ_2$ symmetry (which exchanges the two copies of $SU(2)$). Furthermore one can see from the supergravity equations of motion that the anti-D3 branes and the fields they source do not create a source for the axion $C_0$. So we are looking for a family of non-supersymmetric solutions with  $SU(2) \times SU(2) \times {\mathbb Z}_2$ symmetry which are continuously connected to the KS solution. The most general ansatz consistent with these symmetries was proposed by Papadopoulos and Tseytlin (PT) \cite{Papadopoulos:2000gj}
  \beq
  \label{PTmetric}
 ds_{10}^2=e^{2A+2p-x} ds_{1,3}^2 + e^{-6p-x} d\tau^2 + e^{x+y} (g_1^2+g_2^2) + e^{x-y} (g_3^2+g_4^2) + e^{-6p-x} g_5^2 \ ,
 \eeq
where all functions depend on the variable $\tau$.  The fluxes and dilaton are
 \bea
 \label{PTfluxes}
 H_3&=&\tfrac12 (k-f) g_5 \wedge (g_1 \wedge g_3+ g_2 \wedge g_4) + d\tau \wedge (f' g_1 \wedge g_2+ k' g_3 \wedge g_4) \ , \nn \\
 F_3&=&F g_1 \wedge g_2 \wedge g_5+ (2P-F) \, g_3\wedge g_4 \wedge g_5 +
 F' d\tau \wedge (g_1 \wedge g_3+ g_2 \wedge g_4) \ , \\
 F_5&= &{\cal F}_5 + * {\cal F}_5 \ , \qquad {\cal F}_5= (k F + f(2P -F)) \, g_1 \wedge g_2 \wedge g_3 \wedge g_4 \wedge g_5  \ ,\nn \\
 \Phi&=& \Phi(\tau),\ \ C_0 =0.\non
 \eea
where $P$ is a constant while $f,k$ and $F$ are functions of $
\tau$, and a prime denotes a derivative with respect to $\tau$. There are several different conventions in the literature for the PT ansatz, we compare our conventions to those of others in Appendix \ref{conventions}.

We will denote the set of functions $\phi^a$, $a=1,...,8$, in the following order
\be
\label{phidef}
 \phi^a=(x,y,p,A,f,k,F,\Phi) \ .
 \ee
The metric in the one dimensional Lagrangian  \eq{Lag1} is
\beq
\label{fieldmetric}
G_{ab} \phi^{\prime a} \phi^{\prime b} = e^{4p + 4A} \left(x'^2+ \frac12 y'^2+6p'^2-6 A'^2+
\frac14 e^{-\Phi-2x} (e^{-2y} f'^2+e^{2y} k'^2+ 2 e^{2\Phi} F'^2 )+ \tfrac14 \Phi'^2 \right) \non
\eeq
and the superpotential is
\beq
\label{superpotential}
W(\phi)=e^{4A-2p-2x}+e^{4A+4p} \cosh y + \frac12 e^{4A+4p-2x} \left( f \, (2P-F)+kF \right).
\eeq

At this point it is worth emphasizing an important point regarding the
choice of definition for the radial coordinate $\tau$. If one
redefines $\tau$ in the ten-dimensional ansatz, or equivalently in the
Lagrangian \eq{Lag1}, this will lead to a significant alteration in
the equations of motion \eq{xieq} and \eq{phieq} since the metric
$G_{ab}$ and the connection $M^a_{\ b}$ transform nontrivially.
Moreover this also leads to an alteration in the {\it definition} of
the $\xi_a$ \eq{xidef}. Alternatively one might choose to redefine the
radial coordinate after having derived the equations of motion
\eq{xieq} and \eq{phieq} in which case the only transformation comes
from the derivative with respect to the radial coordinate so
essentially the other terms just get a multiplicative factor of $(\del
\tilde{\tau}/\del \tau)^{-1}$. Of course both such redefinitions are
completely legitimate, however they are not equivalent since the latter
does not alter the definition of the $\xi_a$.

This is an important consideration since a judicious choice of radial
variable can lead to a significant simplification in the equations of
motion.  This must be kept in mind when comparing the radial
coordinates in \cite{Borokhov:2002fm} and \cite{Kuperstein:2003yt}.
The authors of \cite{Kuperstein:2003yt} found a particularly natural
choice such that the equations simplify somewhat and we will employ
the same choice in the current work (see Appendix \ref{conventions} for more details).

 %%%%%%%%%%%%%%%%%%%%%%%%%%%%%%%%%%
 \subsection{The zero-th order solution: Klebanov-Strassler}

 Since we are expanding around the Klebanov-Strassler solution, we tabulate the form of the zeroth-order functions here:
 \bea \label{KSbackground}
 e^{x_0}&=& \frac14 \, h^{1/2}(\tfrac12 \sinh(2 \tau) - \tau)^{1/3} \ , \nn \\
 e^{y_0}&=&\tanh(\tau/2) \ , \nn \\
 e^{6 p_0}&=&  \, \frac{  24 \, (\tfrac12 \sinh(2 \tau) - \tau)^{1/3}}{ h \, \sinh^2\tau}  \ , \nn \\
 e^{6A_0}&=&\frac{1}{3 \cdot 2^9} \, h (\tfrac12 \sinh(2 \tau) - \tau)^{2/3} \sinh^2\tau  \ , \\
 f_0&=&-P\frac{  \, (\tau \coth \tau -1)(\cosh \tau -1)}{\sinh \tau}, \nn\\
 k_0&=&-P \frac{\,(\tau \coth \tau -1)(\cosh \tau +1)}{\sinh \tau}, \nn \\
  F_0&=& P\frac{(\sinh \tau -\tau)}{\sinh \tau} ,\nn \\
  \Phi_0&=&0\nn
 \eea
 where the function $h(\tau)$ related to the warp factor is given by the integral\footnote{We have $h_0=32 P^2 \int_0^{\infty} \frac{\tau \coth \tau -1}{\sinh^2 \tau} (\tfrac12 \sinh(2 \tau) - \tau)^{1/3} d\tau=18.2373 P^2$.}
 \bea
 h&=&e^{-4A_0-4p_0+2x_0} \non \\
 &=&h_0-32 P^2 \int_0^\tau \frac{t \coth t -1}{\sinh^2t} (\tfrac12 \sinh(2 t) - t)^{1/3} dt \ . \label{KSh}
 \eea

There is an important difference between the Klebanov-Tseytlin (KT) solution \cite{Klebanov:2000nc} and the Klebanov-Strassler (KS) one \cite{Klebanov:2000hb}. The KT solution has a warp factor
\be
h_{KT}(r)= \frac{27\pi \Blp N+a P^2 \ln (r/r_0) +a P^2/4 \Brp}{4 r^4},\ \ a=3/(2\pi)
\ee
which has a naked singularity at some finite distance $r_s$ where $h$ becomes zero. The
leading UV behavior of the KT solution does in fact agree with the KS
solution with the identification $r=e^{\tau/3}$ but the subleading behavior of the KT solution is ambiguous
due to ability to absorb a shift into the singularity. For this reason
it is not reasonable to compare the subleading behavior of the KT and
KS solution and in fact deformations of the KT solution (in particular
the deformation proposed in \cite{DeWolfe:2008zy}) which have
subleading terms are not in general solutions of \eq{xieq} and
\eq{phieq} with the $\phi_0^a$ of the KS solution.

%%%%%%%%%%%%%%%%%%%%%%%%%%%%%%%%%%
\subsection{$\xi_i$ Equations}
%%%%%%%%%%%%%%%%%%%%%%%%%%%%%%%%%%

The first step is to solve the system of equations \eq{xieq} for
$\xi_i$. In \cite{Borokhov:2002fm} this system was solved
perturbatively in the UV and IR seperately. Since our goal in this
work is to connect the UV and IR asymptotics, that analysis will not
suffice for our purposes. In \cite{Kuperstein:2003yt} this system was
solved analytically with the assumption that $
\xi_{1}=\xi_{3}=\xi_{4}=0$, which remarkably corresponds exactly to
the non-supersymmetric deformation generated by a mass term for the
gaugino in the dual field theory (often referred to as the $N=0^*$
flow). This solution is still not general enough for our purposes, we
need to lift all of these constraints and consider the most general
solution space.

 We find that to solve the $\xi_i$-equations in general, it is convenient to pass to the basis\footnote{The inverse is
 $\xi_a=\left( \tfrac13(3\txi_1+\txi_3+\txi_4),\txi_2, \txi_1+\txi_3, -\txi_1- \txi_4,\tfrac12(\txi_5+\txi_6),\tfrac12(\txi_5+\txi_6),\txi_7,\txi_8\right)$ }
\be
\txi_a\equiv\left(3\xi_1-\xi_3+\xi_4,\xi_2,-3\xi_1+2\xi_3-\xi_4,-3\xi_1+\xi_3-2\xi_4,\xi_5+\xi_6,\xi_5-\xi_6,\xi_7,\xi_8 \right).
\ee
The equations in the order in which we solve them, are
 \bea
 \txi_1'&=&e^{-2x_0} \left(2P f_0-F_0(f_0-k_0)\right) \txi_1 \label{txi1eq} \\
 \txi_4'&=& -e^{-2x_0} \left(2P f_0-F_0(f_0-k_0)\right) \txi_1  \label{txi4eq} \\
\txi_5'&=&-\frac13 P e^{-2x_0} \txi_1 \label{txi5eq} \\
\txi_6'&=&-\txi_7-\frac13 e^{-2x_0} (P-F_0) \txi_1  \label{txi6eq} \\
\txi_7'&=&-\sinh(2y_0) \txi_5-\cosh(2y_0)\txi_6+\frac16 e^{-2x_0}(f_0-k_0) \txi_1 \label{txi7eq} \\
\txi_8'&=&(P e^{2y_0}-\sinh(2y_0)F_0) \txi_5 +(P e^{2y_0}-\cosh(2y_0)F_0) \txi_6+\frac12 (f_0-k_0) \xi_7 \label{txi8eq} \\
\txi_3'&=&3 e^{-2x_0-6p_0} \txi_3 +  \left(5 e^{-2x_0-6p_0} -e^{2x_0}  (2P f_0-F_0 (f_0-k_0) \right) \txi_1 \label{txi3eq} \\
\txi_2'&=&\txi_2 \cosh y_0 +\frac13 \sinh y_0 (2 \txi_1 +\txi_3 +\txi_4) \non \\
&&\ \ \ \ \ \ \ \ +4\left((P e^{2y_0}-\cosh(2y_0)F_0) \txi_5 + (P e^{2y_0}-\sinh(2y_0)F_0) \txi_6\right)  \label{txi2eq}
   \eea
where a prime indicates derivative with respect to $\tau$.

%%%%%%%%%%%%%%%%%%%%%%%%%%%%%%%%%%
\subsection{$\phi^i$ Equations}
%%%%%%%%%%%%%%%%%%%%%%%%%%%%%%%%%%

Once the $\xi_i$ equations are solved, one can insert the solutions in the $\phi^j$ equations, which we write here explicitly. We start by performing a linear field redefinition\footnote{The inverse is $ \phi^a-\phi^{a}_0=(\frac15 (-2\tphi_1+2\tphi_3+5\tphi_4),\tphi_2,\frac{1}{15} (2\tphi_1+3\tphi_3-5\tphi_4), \frac13 (-\tphi_1+\tphi_4) ,\tphi_5,\tphi_6,\tphi_7,\tphi_8)$ and to avoid confusion we abuse notation at from now on put all indices down.}
 \be
 \tphi_a=(x-2p-5A,y,x+3p, x-2p-2A ,f,k,F,\Phi). \label{tphidef}
 \ee
where all the fields on the right hand side are the first order perturbations.

Then the system of equations is (in the order in which we will solve them)
 \bea
  \tphi_8^{\prime} &=&- 4e^{-4(A_0+p_0)}\txi_8 \label{phi8peq} \\
  \tphi_2^{\prime} &=& - \cosh y_0\, \tphi_2 - 2 e^{-4(A_0+p_0)} \txi_2 \label{phi2peq}\\
 \tphi_3^{\prime} &=&  - 3 e^{-6p_0-2x_0}\tphi_3 - \sinh y_0\, \tphi_2 - \frac16 e^{-4(A_0+p_0)}  (9\txi_1+5\txi_3+2\txi_4)   \label{phi3peq}\\
  \tphi_1^{\prime} &=&  2 e^{-6p_0-2x_0} \tphi_3 -\sinh y_0 \tphi_2 + \frac{1}{6}e^{-4(A_0+p_0)}(\txi_1+3\txi_4)  \label{phi1peq}\\
 \tphi_5^{\prime} &=&   e^{2y_0}(F_0-2P) (2 \tphi_2 + \tphi_8) +e^{2y_0}\tphi_7
 - 2 e^{-4(A_0+p_0)+2(x_0+y_0)} (\txi_5 +\txi_6) \label{phi5peq}\\
 \tphi_6^{\prime}  &=&  e^{-2y_0} (F_0 (2\tphi_2 -\tphi_8)-\tphi_7)  - 2 e^{-4(A_0+p_0)+2(x_0-y_0)}  (\txi_5-\txi_6)  \label{phi6peq}\\
 \tphi_7^{\prime} &=&   \half    \Blp \tphi_5-\tphi_6 +(k_0-f_0)\tphi_8 \Brp -2e^{-4(A_0 + p_0) +2x_0 } \txi_7 \label{phi7peq} \\
  \tphi_4^{\prime} &=& \frac{1}{5} e^{-2x_0} ( f_0 (2P-F_0) +k_0 F_0) (2 \tphi_1-2\tphi_3 -5 \tphi_4)  +\frac{1}{2} e^{-2x_0} (2P-F_0) \tphi_5    \non  \\
 && +\frac{1}{2} e^{-2x_0} F_0  \tphi_6 +\frac{1}{2} e^{-2x_0}  (k_0 -f_0) \tphi_7 -\frac{1}{3} e^{-4(A_0+p_0)}\txi_1 \label{phi4peq}
 \eea

%%%%%%%%%%%%%%%%%%%%%%%%%%%%%%%%%%
\subsection{Imaginary (anti)-Self Duality of The Three Form Flux} \label{sec:selfduality}
%%%%%%%%%%%%%%%%%%%%%%%%%%%%%%%%%%

Before moving on to solving the first order system of equations, it is
worthwhile considering the conditions for self duality of the
three form flux. The complex three-form flux of the KS solution is well known to be
imaginary self dual (as are all warped Calabi-Yau
solutions of IIB supergravity \cite{Gubser:2000vg, Grana:2001xn}) and
one might be tempted to consider the possibility that the flux turned
on by an anti-D3 brane is imaginary anti-self dual. It is important to
note that since the metric is deformed at first order (as opposed to
deformations of $AdS_5$ \cite{Grana:2000jj}), it is not consistent to
consider the self duality of the flux with respect to the background
metric. Instead, one should analyze the condition for self duality
using the perturbed metric.

In our conventions, the condition for imaginary self duality is
\be
{\rm ISD}:\ \ e^{\Phi} *F_3 = -H_3 \ ,
\ee
and the condition for imaginary anti self duality is
\be
{\rm anti-ISD}:\ \ e^{\Phi} *F_3 =+H_3.
\ee
In terms of the PT ansatz, these conditions lead to
\bea
e^\Phi F e^{-2y} &=&\mp k' \non \\
e^\Phi  (2P-F) e^{2y} &=&\mp f' \\
 F'&=& \mp\half e^{-\Phi} (k-f). \non
\eea
with the upper sign for the ISD condition.

Using Eqs (\ref{phi5peq}-\ref{phi8peq}) and the fact that the zeroth order three-form flux is imaginary self dual, the conditions that the first order flux be ISD  become
\bea
(\txi_5-\txi_6) &=&0,\non  \\
(\txi_5+\txi_6) &=&0, \label{ISDeval} \\
 \txi_7&=&0. \non
 \eea
 It is easy to see that these immediately lead to $\txi_1=0$ and force
 $(\txi_4,\txi_8)$ to be constant. Hence, the generic solution to the
 system of first-order equations \cite{Borokhov:2002fm} does not have purely
 ISD flux, contrary to recent claims in \cite{McGuirk:2009xx}. One can
 also see this by examining the small-$\tau$ expansion of the dilaton.
 ISD fluxes do not source the dilaton. Allowing for a delta-function source at the origin, the dilaton in an
 ISD background should be equal to
\be \label{dilgreen}
\phi = C_1 + C_2 \Blp {1\over \tau} + {2 \tau \over 15} - {\tau^3 \over 315} + \ldots\Brp
\ee
where the constant $C_2$ multiples the free Green's function of the
deformed conifold \cite{McGuirk:2009xx}. It turns out that the full
expansion of the dilaton near the origin also contains even powers of
$\tau$. Setting those to zero and demanding the coefficient of the odd
terms to be proportional to those in the Green's function
(\ref{dilgreen}) implies certain conditions on the integration
constants ($X_1=0$, $X_5 = -X_6 = X_7$), which are actually weaker
than (\ref{ISDeval}).

It is interesting to note that all BPS solutions of
(\ref{xieq},\ref{phieq}) have ISD three-form fluxes since for supersymmetric solutions all the $\xi$'s vanish. If one relaxes
the $\ZZ_2$ symmetry in the PT ansatz then there exist BPS solutions
with non-ISD flux \cite{Gubser:2004qj, Butti:2004pk}.

The conditions for the first order deformation to be anti-ISD are of course more involved:
\bea
(\txi_5-\txi_6) &=& e^{4(A_0+p_0)-2x_0} \Blp F_{0}  (2\tphi_2 -\tphi_8)+   \tphi_7 \Brp ,\non \\
(\txi_5 +\txi_6) &=&e^{4(A_0+p_0)-2x_0}  \Blp (F_0-2P) ( 2\tphi^2+\tphi^8 )+ \tphi^7\Brp, \label{aISDeval}\\
\non \\
    \txi_7 &=&-\half e^{4(A_0 + p_0) -2x_0 }\Blp  (\tphi^6-\tphi^5) - \tphi^8(k_0-f_0) \Brp. \non
\eea
One can try arguing that these conditions are necessary in order for
the first-order perturbed solution to correspond to anti-D3 branes.
However, as we will find in the next section there exists a much
simpler and more physical signature of the presence of anti-D3 branes:
the force on a probe D3 brane.

%%%%%%%%%%%%%%%%%%%%%%%%%%%%%%%%%%
\section{The Force on a Probe D3 Brane} \label{forcesection}
%%%%%%%%%%%%%%%%%%%%%%%%%%%%%%%%%%

Before solving the above equations, we compute the force on a probe
D3 brane in the perturbed solutions. As is well known, such a brane
feels no force in the unperturbed Klebanov-Strassler solution. It
appears at first glance that both the Dirac-Born-Infeld (DBI) and the Wess-Zumino (WZ)
actions are rather complicated in the perturbed solution, however we
show below that by using the first order equations of motion, most
terms in the total action cancel and the final expression for the force
on a probe D3 brane is quite simple.

The DBI action for the probe D3 brane  is
\bea
V^{DBI}& =&  \sqrt{- g_{00} g_{11} g_{22} g_{33} } \non \\
&=& e^{4A +4p -2x},
\eea
and in the first-order expansion, the derivative of this action with respect to $\tau$ is
\bea
F^{DBI}&=&- {d V^{DBI} \over d \tau} \non \\
&=& - e^{4A_0 +4p_0 -2x_0}(1+ 4A_1 +4p_1 -2x_1)(4A_0' +4p_0' -2x_0'+4A'_1 +4p'_1 -2x'_1)  \nonumber \\
&= & - {d V_0^{DBI} \over d \tau} - e^{4A_0 +4p_0 -2x_0} (-2 \tilde \phi_4 (4A_0' +4p_0' -2x_0') -2 \tilde \phi'_4)  ~,
\eea
where $V_0^{DBI}$ is the DBI action in the unperturbed solution,
\be
F_{0}^{DBI}= - e^{4A_0 +4p_0 -2x_0}(4A_0' +4p_0' -2x_0') \ ,
\ee
and we have used the definition of $\tilde \phi_4 $ in \eq{tphidef}.

To compute the form of the WZ action one needs to know the value of
the RR potential $C^{(4)}$ along the worldvolume of the brane, which
we obtain by integrating the RR field strength $F^{(5)}$. It is
however easier to compute instead the derivative of this action with
respect to $\tau$, which gives the force exerted on the D3 brane by
the RR field:
\bea
F^{WZ}&=&- {d V^{WZ} \over d \tau} \non \\
&=& F^{(5)}_{0123\tau} \non \\
&=& -(kF+f (2P-F)) e^{4A +4p -4x}. \non \eea
The zero-th order and first order WZ forces are
\bea
F_0^{WZ}&=& -(k_0F_0+f_0 (2P-F_0)) e^{4A_0 +4p_0 -4x_0} \\
F_1^{WZ}&=& - e^{4A_0 +4p_0 -4x_0} \left[{(k_0F_0+f_0 (2P-F_0))(4A_1 +4p_1 -4x_1)} \right] \nonumber \\
&~&- e^{4A_0 +4p_0 -4x_0}\left[{ k_1 F_0 + F_1(k_0-f_0) + f_1 (2P-F_0)
  }\right] \label{fwz}.
\eea
It is not hard to check that the
zeroth-order WZ and DBI contributions to the force cancel, as expected
for a BPS D3 brane in the KS background.

Using this, as well as the equation of motion for $\tilde \phi_4'$ we get
\bea
F^{DBI}_1&=& - e^{4A_0 +4p_0 -2x_0} (-2 \tilde \phi_4 (4A_0' +4p_0' -2x_0') -2 \tilde \phi'_4)\nonumber \\
&=& 2 e^{4A_0 +4p_0 -4x_0} \left[{{1\over 5}(k_0F_0+f_0 (2P-F_0)) (2 \tilde \phi_1-2 \tilde \phi_3 - 10 \tilde \phi_4)}\right] \nonumber \\
&~&+    e^{4A_0 +4p_0 -4x_0} \left[{k_1 F_0 + F_1(k_0-f_0) + f_1 (2P-F_0) }\right] + {2\over 3} e^{-2x_0} \tilde\xi_1~.
\label{fbi}
\eea
From the definition of the $\tphi^i$, we have that
\be
2 \tilde \phi_1-2 \tilde \phi_3 - 10 \tilde \phi_4= 10 (A_1 + p_1 -x_1),\non
\ee
thus the first and the second term of the Wess-Zumino force (\ref{fwz}) cancel against corresponding terms from the DBI force (\ref{fbi}). Hence to first order in perturbation theory, the only contribution to the force felt by a probe D3 brane comes from the term proportional to $\tilde\xi_1$:
\bea
F=F^{DBI} + F^{WZ}=  {2\over 3} e^{-2x_0} \tilde\xi_1 .
\eea

As we will show in detail in section \ref{sec:solution}, it is rather
straightforward to compute the UV asymptotic expansion of $
\tilde\xi_1 $ by expanding the integrand in (\ref{txi1}) for large
values of $\tau$:
\be
\tilde\xi_1 = X_1 3 (1-4\tau) e^{-4 \tau/3} + {\cal O}(e^{-10\tau/3})
\ee
By expanding in the ultraviolet $ e^{-2x_0}$ as well
\be
e^{-2x_0} = {8 \over 3 P^2 (4 \tau -1)} + {\cal O}(e^{-2\tau})  \ ,
\ee
we can see that the UV expansion of the force felt by a probe D3 brane in the
first-order perturbed solution is always
\be
F_\tau  \sim X_1 e^{-4 \tau/3} + {\cal O}(e^{-10\tau/3})  \ .
\ee
Recalling that in the UV, $\tau$ is related to the canonical radial coordinate
\beq \label{rtau}
r=e^{\tau/3} \ ,
\eeq
it follows that
\beq \label{forcer}
F_r \sim \frac{X_1}{r^5} +  {\cal O}\left(\frac{1}{r^{11}}\right)\
\eeq
and thus the potential goes like
\beq
V \sim C+ \frac{X_1}{r^4} \ .
\eeq
Having obtained this simple universal result, we pause to consider its
physical implications.  First, the force is largely independent of
which modes perturb the KS solution
and on whether these modes are
normalizable or non-normalizable; it may be zero in certain solutions
(such as \cite{Kuperstein:2003yt}, where $X_1=0$) but when non-zero it has a
universal form (\ref{forcer}).  Out of the 14 physically-relevant $SU(2) \times SU(2)
\times {\mathbb Z}_2$-preserving perturbation parameters, only one
enters in the force.

Second, this result agrees with that obtained in
\cite{Kachru:2003sx}.  In that paper, the
force felt by a probe anti-D3 brane in KS with D3 branes at the
bottom was shown to scale like $1/r^5$ and to be linear in
the D3 charge. It was also argued using Newton's third law that
the same force should be felt by a probe D3-brane in KS with anti-D3 branes
at the bottom.
Our result implies therefore that in order to describe anti-D3 branes in the infrared, one {\it must}
have a nontrivial $\txi_1$, and furthermore that the constant
$X_1$ must be proportional to the number of anti-D3 branes.

We would also like to note that in the ultraviolet the mode (in the $\phi_i$)
proportional to $X_1$ is a normalizable mode decaying as $r^{-8}$, in
fact this is the most convergent of the 14 modes. The
necessity for having such a mode in order to find the $ r^{-5}$ force
predicted by \cite{Kachru:2003sx} was discussed in
\cite{DeWolfe:2008zy}, although the mode itself was not identified.
 However, the force analysis
alone in no way supports or disfavors the possibility that an
anti-D3 brane sources a non-normalizable mode. The force has exactly
the same $r$-dependence regardless of whether non-normalizable modes
are turned on or not. Hence, the cancelation of terms in the force up to order $1/r^4$
is universal.

%%%%%%%%%%%%%%%%%%%%%%%%%%%%%%%%%%
 \section{The Space of Solutions}    \label{sec:solution}
%%%%%%%%%%%%%%%%%%%%%%%%%%%%%%%%%%

In this section we find the generic solution to the system
(\ref{txi1eq}-\ref{phi4peq}), which depends on sixteen integration
constants (although only fourteen are physical). As we will see, the full solution involves integrals that cannot be done explicitly but these integrals can be expanded both in the UV and
IR, giving rise to sixteen UV and sixteen IR integration constants. Having the expression for the full solution permits one to relate numerically the UV and IR integration constants.

The first equation \eq{txi1eq} is solved by
\beq \label{txi1}
\txi_1= X_1 \exp \left(\int_{0}^\tau d\tau' e^{-2x_0} \left[2P f_0-F_0(f_0-k_0)\right] \right)
\eeq
and this integral cannot be done explicitly.  In a cruel twist, having
argued above that the integration constant $X_1$ is the only mode
which contributes to the force on a D3 brane and is thus the signature
of the anti-D3 brane background, we have immediately found that as long as
$X_1$ is nonzero there is no analytic expression for any of the $\xi_i$
modes. On the other hand, when $X_1=0$, we can fully solve the system
of equations for the $\xi_i$ analytically.  We proceed by initially
setting $X_1$ to zero, finding an analytic expression for all the
other $\xi_i$, and then building on top of this, the integral
expressions for the full solution with nonzero $X_1$.

%%%%%%%%%%%%%%%%%%%%%%%%%%%%%%%%%%
\subsection{Solving the $\xi_i$ Equations for $X_1=0$} \label{solvingxi}

Here we solve the $\xi_i$ equations for the choice
\be
X_1=0 \Rightarrow \txi_1=0
\ee
and later we will find integral expressions for the full solution including $\txi_1$. Having set $\txi_1=0$, we then immediately read off that $(\txi_4,\txi_5)$ are constant
\be
\txi_4= X_4,\ \ \txi_5= X_5.
\ee
 We now have a pair of coupled o.d.e's for $(\txi_6,\txi_7)$
 \bea
\txi_6'&=&-\txi_7  \label{txi67one}, \\
\txi_7'&=&-\sinh(2y_0) X_5-\cosh(2y_0)\txi_6. \label{txi67two}
 \eea
Doing the derivative of (\ref{txi67one}) and using (\ref{txi67two}) one easily determines the solution
 \bea
 \txi_6&=& \frac{(1+2\tau) X_5 + X_6 - 2\tau X_7 + X_7 \sinh (2\tau)}{2\sinh \tau},  \\
 \txi_7&=& \frac{ ((1+ 2\tau )X_5 +X_6-2\tau X_7)\cosh \tau - (2X_5 +X_7 (\cosh\, 2\tau-3))\sinh\tau}{2\sinh^2 \tau}.
 \eea
To ease notation the two integration constants appearing in the second order equation for $\txi_6$ have been denoted
$X_6$ and $X_7$, but one should keep in mind that they both appear in $\txi_6$ and $\txi_7$.
 To compare to the solution in \cite{Kuperstein:2003yt} one sets $X_7=-X_6=X_5$.

 We then directly integrate \eq{txi8eq} and \eq{txi3eq} to find
  \bea
\label{txi8}
\txi_8&=& (1-e^{2\tau})^{-3} \Blp 2 e^{3\tau} \blp P ((-1 + 2 \tau + 4 \tau^2) X_5 +X_7+2\tau (X_6-2 \tau X_7)) \cosh \tau \non \\
&&+       P (X_5 - X_7) \cosh\, 3\tau + (3 X_8 +
       P (X_5 (1- 5 \tau )- 2 X_6 + (-3 + 5 \tau) X_7)) \sinh
      \tau \non \\
      && - (X_8 + P (1 + \tau) (X_5 - X_7)) \sinh\, 3 \tau \brp \Brp,  \\
\label{txi3} \txi_3&=&2 X_3 \blp  \sinh\, 2\tau -2\tau  \brp .
 \eea
Finally we insert the result for $\txi_3$ into \eq{txi2eq} and solve the o.d.e. to find
\bea \label{txi2}
\txi_2&=&\frac16  \Blp (2 X_4+ 4 \tau (-4 X_3+3 P X_7) ) \cosh \tau  + 2 (6 X_2-X_4) \sinh \tau  \Brp \non \\
&& + \frac{P \cosh \tau}{2 \sinh^2 \tau} \blp(2 (X_5 - X_7)-(X_5 (1+ 4 \tau) + X_6 - 4 \tau X_7 \brp \nn \\
&& + \frac{P \tau}{2 \sinh^3 \tau} (X_5 (1+ 2 \tau) + X_6 - 2 \tau X_7).
\eea

So far we have introduced $8$ integration constants
\be
(X_1,X_2,X_3,X_4,X_5,X_6,X_7,X_8)
\ee
however the the zero-energy condition is a constraint amongst these
\be \label{zeroenergy}
-X_4+6X_2-3P X_5 - 4 X_3+ 9P X_7 =0 \ .
\ee
This leaves $7$ independent constants, but for the time being we will keep the 8 constants since this will provide a clearer picture of the UV asymptotics.

%%%%%%%%%%%%%%%%%%%%%%%%%%%%%%%%%%
\subsection{Solution for $X_1 \neq 0$, UV and IR expansions} \label{sec:X1eq0}

When $\txi_1$ is nonzero, one can find integral expressions for the
solution, but the integrals cannot be performed explicitly since
$\txi_1$ itself, given in (\ref{txi1}), cannot be found explicitly.
$\txi_1$ appears on the right hand side in all of the $\txi$ equations
(except for the equation for $\txi_8$, but the right hand side of
(\ref{txi8eq}) contains $\txi_{5,6,7}$ which are modified by $\txi_1$)
and therefore all $\txi$'s get an extra contribution proportional to
$X_1$. A simple way to understand this extra contribution, is to
recall the following method to solve a system of coupled o.d.e's,
which will in fact be used to solve the equations for $\phi$. Given
the system
\be
g'_i = A_i^{\ j} g_j + b_i
\ee
where $i=1,\ldots,n$, and $g, A, b$ are functions of a single variable $\tau$, one first finds the solutions $g^j_{h,i}$ of the homogeneous equations (again $j=1,\ldots,n$). Then one constructs the ansatz for the inhomogeneous solution by a linear combination of the homogeneous solutions where the coefficients are promoted to functions, and then it is easy to see that one has
\be \label{soltosysode}
g_i=\sum_j g_{h,i}^j  \lam_j (\tau)   \ , \qquad  \lam_j=\int (g_h^{-1})_j{}^i b_i.
\ee
The integration constant here corresponds to adding any amount of the homogeneous solution to $g_i$.
For the simple case $n=1$, that will appear repeatedly in the following, we have
\beq
\label{soltoode}
g(\tau)=\lambda(\tau) g_h(\tau) \ , \quad g_h=e^{\int A}  \ , \quad \lambda=\int_{\tau_0}^\tau \frac{b(\tau')}{g_h(\tau')}  d\tau' \ ,
\eeq
where the integration constant is related to the choice of $\tau_0$.

Adding $\txi_1$ on the right hand side of (\ref{txi4eq}-\ref{txi2eq}) amounts to the change $b_i \to b_i + X_1 f_i(\tau)
\equiv b_i^0+X_1 b_i^1$ where $b_i^1$ contains $\xi_1$ \footnote{The indices 0 and 1 in $b$ are used to distinguish between  the expression without $\txi_1$ and that purely due to $\txi_1$. They should not be confused with zero and first order in perturbation theory (they are both first order).}. This implies
\beq \label{xi1contrlambda}
\lambda_i \to \lambda_i + X_1 \int_0^\tau d \tau' (g_h^{-1})_{i}{}^j b^1_j\equiv \lambda_i^0 + X_1\,  \lambda_i^{1} \ , \qquad \
g_i \to g_i + X_1\, g_{h,i}^j \, \lam_j^{1} \ .
\eeq
Note that in the expression for $\lambda$ we have chosen to absorb the $n$ overall integration constants in $\lambda_i^0$ and fixed the limits of the integrand in $\lambda_i^1$.

The integrands can always be expanded around the IR and UV, and the integrals can be performed in these limits. Expanding (\ref{txi1}) for small $\tau$ amounts to expanding the integrand in a series of powers of $\tau$, doing the integral explicitly and then expanding the exponential again\footnote{In general, if for a given expression to be integrated in the IR there are negative powers of $\tau$ showing up in the expansion, we subtract the infinite contribution from the lower limit of the integral, namely we replace $\int_0^\tau f_n t^{-n} \to \int^\tau f_n t^{-n} = f_n \frac{t^{-n+1}}{-n+1}$ (for $n\neq 1$). }. This gives
\beq
\txi_1^{IR}  = X_1^{IR} \left(1-\frac{16 \left(\frac{2}{3}\right)^{1/3} P^2 \, \tau^2}{3 \, h_0}+{\cal O}(\tau^4)\right) \ .
\eeq
The contribution from $\txi_1$ on the rest of the $\txi$'s can be obtained similarly by expanding the integrands in $\lambda^a_1$.
We give the IR behavior of all the $\xi$'s in Appendix \ref{app:IRxi}.

The UV behavior is slightly more tricky. On one hand, the expansion is in powers of $e^{-\tau/3}\sim 1/r$, where at each power
there are terms proportional to $\tau^n\sim (\log r)^n$ (typically only $n=0$ and $n=1$ appear). Fortunately all these terms can easily be integrated.
On the other hand, expressions like (\ref{txi1}) are split into
\bea \label{X1UVIR}
\txi_1&=&X_1 \exp  \left(\int_{0}^\tau d\tau' f(\tau') \right) = X_1^{IR} \exp  \left(\int_{0}^\infty d\tau' f(\tau') \right) \exp  \left( \int_{\infty}^\tau d\tau' f(\tau') \right)     \nn \\
&\equiv& X_1^{UV} \exp  \left( \int_{\infty}^\tau d\tau' f(\tau') \right)
\eea
where the integrand can now be expanded around large $\tau$, and the integral can be done analytically. The integral from zero to infinity giving the relation between $X_1^{UV}$ and $X_1\equiv X_1^{IR}$ can in principle be done numerically.  This gives
\beq
\txi_1^{UV}=X_1^{UV} e^{-4 \tau /3} (3 - 12 \tau) + {\cal O}(e^{-10 \tau/3})
\eeq

For the other $\txi$'s, one uses the same expansions to obtain the UV behavior of $\lambda^1_i$. We get from (\ref{xi1contrlambda})
\beq
X_1 \, \lam^1_i= X_1 \left(\int_0^\infty  d \tau' (g_h^{-1})_{i}{}^j b^1_j + \int_\infty^\tau d \tau'  (g_h^{-1})_{i}{}^j b^1_j \right) = X_1
\left( k_i +\int_\infty^\tau d \tau'  (g_h^{-1})_{i}{}^j b^1_j \right) \ ,
\eeq
where $k_i$ are some constants that can be obtained numerically, and in the remaining integral one expands the integrand for large $\tau$ and performs the integral analytically. The contributions $X_1 k_i$ shift the constants of integration in $\lambda^0_i$. We therefore define
\beq
X_a^{UV} \equiv X_a^{IR} + X_1 k_a \ .
\eeq
We give the UV expansion of all the $\xi$'s in Appendix \ref{App:UVxi}.

%%%%%%%%%%%%%%%%%%%%%%%%%%%%%%%%%%
\subsection{Solving the $\phi^i$ Equations} \label{solvingphi}

%%%%%%%%%%%%%%%%%%%%%%%%%%%%%%%%%%
\subsubsection{The Space of Solutions}  \label{app:phisol}
%%%%%%%%%%%%%%%%%%%%%%%%%%%%%%%%%%

Here we show how to solve the system of $\phi^i$ equations  (\ref{phi8peq}-\ref{phi4peq}), assuming
$X_1=0$. First we note that $\tphi_8$ is given by the following integral, which  however cannot be solved analytically
\beq
\tphi_8= -64 \int d\tau \frac{\txi_8}{(\tfrac12 \sinh(2 \tau) - \tau)^{2/3}} + Y_8^{IR},
\eeq
where $\txi_8$ is given in (\ref{txi8}).

The equation for $\tphi^2$ is solved using (\ref{soltoode}), which implies
\begin{equation} \label{solphi2}
\tilde \phi_2= \frac{\tilde \lambda^2(\tau)}{\sinh \tau} \ , \quad \tilde \lambda^2=-32 \int d\tau \frac{\sinh \tau\, \txi_2}{(\tfrac12 \sinh(2 \tau) - \tau)^{2/3}} + Y_2^{IR} \,
\end{equation}
where $(\sinh \tau)^{-1}$ is the solution to the homogeneous equation, and $\txi_2$ is given in (\ref{txi2}).
The same trick is used for $\tphi^3$, which is given by
 \bea \label{solphi3}
\tilde \phi_3&=& \frac{\tilde \lambda_3(\tau)}{\sinh (2\tau)-2\tau }\ ,  \\
 \tilde \lambda_3&=&\int d\tau (\sinh (2\tau)-2\tau)\left( \frac{\tphi_2}{\sinh\tau}-\frac{16}{3} \frac{ (\sinh (2\tau)-2\tau ) 5 \tix_3+  \tix_4 }{(\tfrac12 \sinh(2 \tau) - \tau)^{2/3}}\right) + Y_3^{IR} \, . \non
\eea
We obtain an integral expression for $\tphi^1$:
\beq \label{solphi1}
\tphi_1= \int d\tau \left(\frac83 \frac{\sinh^2 \tau \tphi_3}{\sinh(2 \tau) - 2 \tau}+ \frac{\tphi_2}{\sinh \tau}+\frac{32}{9}  \frac{  \tix_4 }{(\tfrac12 \sinh(2 \tau) - \tau)^{2/3}} \right)  + Y_1^{IR}\ .
\eeq
and then using (\ref{soltosysode}), we find that the fluxes $(\tphi_5, \tphi_6,\tphi_7)=(f,k,F)$ are
given by
\beq \label{phi567}
\begin{pmatrix}
\tphi_5 \\ \tphi_6 \\ \tphi_7
\end{pmatrix} = \begin{pmatrix} 2\tau-\sinh\tau -\tanh(\tau/2)-\frac{\tau}{2 \cosh^2(\tau/2)} & \frac{1}{2 \cosh^2(\tau/2)} & 1\\
2\tau+\sinh\tau -\coth(\tau/2)+\frac{\tau}{2 \sinh^2(\tau/2)} & -\frac{1}{2 \sinh^2(\tau/2)} & 1\\
\frac{\tau}{\sinh\tau}-\cosh\tau &-\frac{1}{\sinh\tau}& 0 \end{pmatrix} \begin{pmatrix}
\lambda_{5} \\ \lambda_{6} \\ \lambda_{7}  \end{pmatrix} \ .
\eeq
This $3 \times 3$ matrix contains the solutions to the homogeneous equations, and the derivatives of the functions $\lambda_a$
are given by
\beq
\begin{pmatrix}
\lambda_5^{'} \\ \lambda_6^{'} \\ \lambda_7^{'}
\end{pmatrix}
= \begin{pmatrix} - \frac{1}{4 \sinh\tau} &  \frac{1}{4 \sinh\tau} & - \frac{\cosh\tau}{2 \sinh^2\tau} \\
\frac14 (\cosh\tau-\frac{\tau}{\sinh\tau}) & -\frac14 (\cosh\tau-\frac{\tau}{\sinh\tau}) & -\frac{(\sinh(3\tau)+4 \tau \cosh\tau -7 \sinh\tau)}{8 \sinh^2\tau}
 \\ \frac12 (1+\frac{\tau}{\sinh\tau}) & \frac12 (1-\frac{\tau}{\sinh\tau})
& \frac{1}{\sinh\tau}(-1+\tau \frac{\cosh\tau}{\sinh\tau}) \end{pmatrix} \begin{pmatrix} b_5 \\ b_6\\b_7
\end{pmatrix}
\eeq
where $b_5, b_6, b_7$ are the rhs of (\ref{phi5peq}), (\ref{phi6peq}) and (\ref{phi7peq}) respectively, and the $3 \times 3 $ matrix is the inverse of the one in (\ref{phi567}). We will call the constants of integration in these equations $Y_5, Y_6, Y_7$, even though the three functions depend on the three of them (i.e. the integration constant in $\tphi_5$ is not just $Y_5$, but a combination of $Y_{5,6,7}$).

Finally, the equations for $\tphi^4$ is solved using the same trick, namely (\ref{soltoode})
except that the solution to the homogeneous equation can be found only as an integral expression.
Formally, it is
\beq
\tphi_4=\lambda_4 \tphi_{4h} \  \qquad , \ \lambda_4=\int \frac{b_4}{\tphi_{4h}}+Y_4^{IR}
\eeq
where $b_4$ is the right hand side of (\ref{phi4peq}) (setting $\tphi^4$ to zero) and
\beq
\tphi_{4h}={\rm exp}\left[8\, P^2 \int d\tau \frac{\tau \sinh(3\tau)-\cosh(3\tau)+5\tau \sinh\tau - \cosh\tau (4\tau^2-1) }{h(\tau) \sinh^3\tau (\cosh\tau \sinh\tau-\tau)^{2/3}}\right]\ .
\eeq

 \subsubsection{IR behavior} \label{sec:phiIR}
 We now give the IR expansions
 of the $\phi^i$. We write the divergent terms and the constant
 terms since terms which are regular in the IR will not provide any
 constraints on our solution space. Here we have used the zero energy condition 
 \eq{zeroec} to substitute for $X_2^{IR}$.

\bea
\tphi_8&=&  \frac{16}{\tau}(2/3)^{1/3}   (6 X_8^{IR} + P (5 X_5^{IR} - X_6^{IR}- 6 X_7^{IR}))+ Y_8^{IR}  +\cO(\tau)\non  \\
\tphi_2&=&\!\!\!\!
\frac{1}{\tau} Y_2^{IR} \!+\! \frac{\log \tau}{\tau} 16(2/3)^{1/3} \Blp \!\!-\!\!\blp \frac{1}{3}  \!+\! 16 (2/3)^{1/3} \frac{P^2}{h_0}\brp X_1^{IR}  \!+\! (-2X_4^{IR} \!+\! P X_5^{IR} \!+\! P X_6^{IR}) \Brp +\cO(\tau)\non \\
\tphi_3&=&
\frac{1}{\tau^3}\frac{3 Y_3^{IR}}{4} + \frac{1}{\tau}\Blp \frac{Y_2^{IR}}{2}-\frac{3Y_3^{IR}}{20} + \frac{X_1^{IR}64 (2/3)^{2/3}P^2 }{h_0}-P(X_5^{IR} +X_6^{IR})4 (2/3)^{1/3} \Brp  \nn \\
&& \! \! \! \! \! \!  \! \! \! \! \! \! \! \! \! \! \! \!
+\frac{\log \tau}{\tau} \Blp\!\! -\!\! X_1^{IR}\blp 8(2/3)^{1/3}/3 + (2/3)^{2/3}128 P^2/h_0\brp +8 (2/3)^{1/3}(- 2 X_4^{IR}+P(X_5^{IR}+X_6)) \Brp  +\cO(\tau) \non\\
\tphi_1&=&- \frac{1}{\tau^3} \frac{Y_3^{IR}}{2} + \frac{1}{\tau} \Blp -2Y_2^{IR} + \frac{Y_3^{IR}}{10}+ X_1^{IR}\Blp 10 (2/3)^{4/3} + \frac{2^{2/3} 3^{1/3}128  P^2}{h_0} \Brp \non \\
&& + 4(2/3)^{1/3} (10X_4^{IR}-6P(X_5^{IR}+X_6^{IR}))\Brp + \frac{\log \, \tau}{\tau} \Blp X_1^{IR}\Blp 16 (2/3)^{4/3} + \frac{(2/3)^{2/3} 512  P^2}{h_0} \Brp \non \\
&& +X_4^{IR} 64(2/3)^{1/3} - (X_5^{IR}+X_6^{IR}32(2/3)^{1/3})  \Brp + Y_1^{IR} +\cO(\tau) \non \\
\tphi_5&=& \frac{Y_6^{IR}}{2} + Y_7^{IR} +X_1^{IR} \frac{16 P}{3}\Blp \frac{16 (2/3)^{2/3} P^2 }{ h_0}-(2/3)^{1/3}  \Brp  -X_4^{IR} 16(2/3)^{1/3} P+\cO(\tau)
\non \\
\tphi_6&=& \frac{1}{\tau^2} \Blp -2 Y_6^{IR} +  (2/3)^{1/3}32 P X_1^{IR} - (4/3) h_0 (X_5^{IR} + X_6^{IR}) \Brp  \non \\
&& +\Blp  \frac{Y_6^{IR}}{6} + Y_7^{IR} + X_4^{IR} (2/3)^{1/3} 16 P - \Blp \frac{2h_0}{9}  +\frac{64}{3}(1/3)^{1/3}P^2 \Brp X_{5}^{IR} -32 (2/3)^{1/3} P X_8^{IR} \non \\
&&- \frac{2P}{3} Y_{2}^{IR}- \Blp \frac{2h_0}{9}  +\frac{32}{3}(1/3)^{1/3}P^2 \Brp X_{6}^{IR} +32 (2/3)^{1/3}P^2 (X_{6}^{IR}/3 + X_{7}^{IR})\Brp  \non \\
&&+\log \tau \Blp X^{IR}_1 \blp \frac{(2/3)^{1/3} 32 P }{9}+\frac{512(2/3)^{2/3} P^3}{3h_0}\brp + \frac{32 (2/3)^{1/3}P}{3} (2 X_4^{IR}-P^2(X_5^{IR}+X_6^{IR}))   \Brp \non \\
&&+\cO(\tau) \non \\
\tphi_7&=& \frac{1}{\tau} \Blp - Y_6^{IR} + \frac{h_0}{3} (X_5^{IR} + X_{6}^{IR}) \Brp +\cO(\tau)\non \\
\tphi_4&=&  \frac{1}{\tau} \Blp
\!\! -\!\! \frac{16(2/3)^{1/3} P Y_6^{IR} }{h_0}
\!\! -\!\! \frac{2^{1/3} 3^{2/3} 8P Y_7^{IR} }{h_0}  +X_1^{IR} \Blp 8 (2/3)^{1/3} + 64 (2/3)^{2/3} P^2 /h_0 - \frac{4096 P^4}{3 h_0^2} \Brp\non \\
&&+X_4^{IR}\frac{256 (3/2)^{1/3}P^2}{h_0}+8 (2/3)^{4/3} (X_5^{IR} + X_6^{IR}) - \frac{16 (2/3)^{1/3} P^2}{ 3 h_0} Y_3^{IR}
\Brp  + Y_4^{IR} +\cO(\tau) \non
\eea

%%%%%%%%%%%%%%%%%%%%%%%%%%%%%%%%%%
\subsubsection{UV behavior} \label{sec:phiUV}

Here we provide the UV asymptotics for the $\tphi^i$ including terms
which fall off as $e^{-4\tau/3}$ and slower. However as shown in the
table \ref{UVmodetable}, certain modes have leading behavior in the
UV which is more convergent than this.

\bea
\tphi_8&=&  Y_8^{UV}+2^{1/3} 24e^{-4\tau/3} (P (3 + 4 \tau) (X_5^{UV} -X_7^{UV})+ 4 X_8^{UV} ) +\cO(e^{-10\tau/3}) \non \\
\tphi_2&=&-2^{1/3}48 e^{-\tau/3} \Blp 2 X_2^{UV}  - (-3 + 2 \tau) (2 X^{UV}_3/3 - P X_7^{UV} )\Brp +2 e^{-t} Y_2^{UV}+{\cal O}( e^{-7 \tau/3}) \non \\
\tphi_3&=&\!\!\!\!\!\! - 2^{1/3}10  X^{UV}_3 e^{2\tau/3} - e^{-4\tau/3}2^{10/3} \Blp36 X^{UV}_2 + 2  X^{UV}_4 + X^{UV}_3(112 -
  \! \! \frac{137}{3}  \tau) - 36 P X^{UV}_7 (3-\tau)\Brp + {\cal O}( e^{-2 \tau}) \non \\
\tphi_1&=& Y_1^{UV} -2^{1/3}20 X_3^{UV}e^{2\tau/3} + 2^{1/3}4\Blp 108 X_2^{UV} + X_4^{UV} + 176 X_3^{UV} - 189 P X_7^{UV}  \Brp e^{-4\tau/3}    \non \\
&& - e^{-4\tau/3} \tau \,2^{1/3} 16 \Blp \frac{79}{3} X_3^{UV} - 27 P X_7^{UV}  \Brp  + {\cal O}( e^{-2 \tau}) \non \\
\tphi_5&=& -\frac{Y_5^{UV}}{2} e^{\tau}+ (-Y_5^{UV} + Y_7^{UV} ) + (2 Y_{5}^{UV} - P Y_8^{UV}) \tau \non \\
&& +2^{1/3}36 \Blp 2 P X_2^{UV} +7 P X_3^{UV} + 7 P^2 X_7^{UV}   \Brp e^{-\tau/3} - \frac{144 P X_3^{UV}}{3}  e^{-\tau/3} \tau \non \\
&& +\half \Blp 5 Y_5^{UV} + 4 Y_6^{UV} + 2P Y_2^{UV} -2PY_{8}^{UV} \Brp e^{-\tau} + \half \Blp -4 Y_5^{UV} +2P Y_8^{UV}  \Brp e^{-\tau} \tau \non \\
&& +2^{1/3}72 P \Blp -4 X_2^{UV} + X_8^{UV} - 5 X_3^{UV}  +2P (X_5^{UV}+X_6^{UV}) \Brp e^{-4\tau/3}  \non \\
&& -2^{1/3} 72 P\Blp 4 X_2^{UV} - 2P X_5^{UV} -\frac{2X_3^{UV}}{3}+P X_7^{UV} \Brp e^{-4\tau/3} \tau \non \\
&&- 2^{1/3} 72 P\Blp -\frac{8X_3^{UV}}{3}+4P X_7^{UV}\Brp  e^{-4\tau/3} \tau^2 + {\cal O}( e^{-2 \tau}) \non \\
\tphi_6&=& +\frac{Y_5^{UV}}{2} e^{\tau}+ (-Y_5^{UV} + Y_7^{UV} ) + (2 Y_{5}^{UV} - P Y_8^{UV}) \tau \non \\
&& -2^{1/3}36 \Blp 2 P X_2^{UV} +7 P X_3^{UV} + 7 P^2 X_7^{UV}   \Brp e^{-\tau/3} + \frac{144 P X_3^{UV}}{3}  e^{-\tau/3} \tau \non \\
&& -\half \Blp 5 Y_5^{UV} + 4 Y_6^{UV} + 2P Y_2^{UV} -2PY_{8}^{UV} \Brp e^{-\tau} - \half \Blp -4 Y_5^{UV} +2P Y_8^{UV}  \Brp e^{-\tau} \tau \non \\
&& +2^{1/3}72 P \Blp -4 X_2^{UV} + X_8^{UV} - 5 X_3^{UV}  +2P (X_5^{UV}+X_6^{UV}) \Brp e^{-4\tau/3}  \non \\
&& -2^{1/3} 72 P\Blp 4 X_2^{UV} - 2P X_5^{UV} -\frac{2X_3^{UV}}{3}+P X_7^{UV} \Brp e^{-4\tau/3} \tau \non \\
&&- 2^{1/3} 72 P\Blp -\frac{8X_3^{UV}}{3}+4P X_7^{UV}\Brp  e^{-4\tau/3} \tau^2 + {\cal O}( e^{-2 \tau}) \non \\
\tphi_7&=&  -\frac{Y_5^{UV}}{2} e^{\tau}  + 2^{1/3}36 P \Blp -6 X_2^{UV} - (9-4\tau) X_3^{UV} + (10-4\tau)P X_7^{UV}  \Brp e^{-\tau/3}   \non \\
&&+ \Blp  \half(4\tau-1)Y_5^{UV} - 2Y_6^{UV} + P Y_2 + P (\tau-1) Y_8^{UV} \Brp e^{-\tau} + {\cal O}( e^{-7 \tau/3}) \non
\eea
\bea
\tphi_4&=& \frac{Y_4^{UV}}{3(4\tau -1)}e^{4\tau/3}   + \frac{2^{1/3}16 X_3^{UV}(  2 \tau+1)}{(  4 \tau-1)} e^{2\tau/3} \non  \\
&&+ \frac{- 4 Y_7^{UV} + (3+4\tau)Y_8^{UV}}{2P(4 \tau-1)} - \frac{1}{P}Y_5^{UV}-\frac{2Y_1^{UV}}{5}  + \frac{32Y4^{UV}(12-85 \tau +25 \tau^2)}{1125 (1-4\tau)^2}e^{-2\tau/3} \non \\
&&  + \frac{2^{1/3}}{4\tau -1}\Blp  -18P (11+8 \tau) X_5^{UV}  + 72 (1+8\tau)X_2^{UV} + 2(-5+8\tau) X_4^{UV} -72 X_8^{UV} \non \\
&& +9P(95-56 \tau +80 \tau^2)X_7^{UV} -\frac{64 X_3 (-3279+17012 \tau - 20785 \tau^2 + 14900 \tau^3)}{375(4\tau-1)}  \Brp e^{-4\tau/3} + {\cal O}( e^{-2 \tau}) \non
\eea

To understand the holographic physics of the $\tphi^i$ modes it is useful to tabulate the leading UV behavior coming from each mode. For each local operator $\cO_i$ of quantum dimension $\Delta$ in the field theory, the well known holographic dictionary \cite{Balasubramanian:1998de,Banks:1998dd}
 relates two modes in dual $AdS$ space, one normalizable and one non-normalizable. These two gravity modes are dual respectively to the vacuum expectation value (VEV) $\langle 0|\cO_i|0\rangle$ and the deformation of the action $\delta S\sim \int d^{d}x \cO_i$:
\bea
{\rm normalizable\ modes}\ \sim r^{-\Delta}&\leftrightarrow& {\rm field\ theory\ VEV's} \non \\
{\rm non-normalizable\ modes}\ \sim r^{\Delta-4}&\leftrightarrow& {\rm field\ theory\ deformations\  of\  the\ action}, \non
\eea
and we recall that we have in the UV $r =e^{\tau/3}$.
For backgrounds like Klebanov-Strassler which are asymptotically $AdS$ only up to certain logarithm terms, it is expected that this dictionary still holds. In Table 1 we have summarized which integration constants correspond to normalizable and non-normalizable modes. It is very interesting to note that in all cases a normalizable/non-normalizable pair consists of one BPS mode and one non-BPS mode.

 %%%%%%%%%%%%%%%%%%%%%%%%%%%%%%%%%%
\begin{table}[h]
\label{UVmodetable}
\onelinecaptionsfalse
\captionstyle{center}\caption{The UV behavior of all sixteen modes \protect\\ in the  $SU(2)\times SU(2)\times \ZZ_2$ symmetric \protect\\ deformation of the Klebanov-Strassler solution.}
\begin{center}
\begin{tabular}{|c|c|c|c|c|c} \hline
dim $\Delta$ & non-norm/norm & int. constant \\ \hline
8 & $r^4/r^{-8}$ & $Y_4/X_1$  \\\hline
7 & $r^3/r^{-7}$ & $Y_5/X_6$ \\\hline
6 & $r^2/r^{-6}$ & $X_3/Y_3$ \\\hline
5 & $r/r^{-5}$ & $---$ \\\hline
4 & $r^0/r^{-4}$ & $Y_7,Y_8,Y_1/X_5,X_4,X_8$ \\\hline
3 & $r^{-1}/r^{-3}$ & $X_2,X_7/Y_6,Y_2$ \\\hline
2 & $r^{-2}/r^{-2}$ & $---$ \\\hline
\end{tabular}
\end{center}
\end{table}
 %%%%%%%%%%%%%%%%%%%%%%%%%%%%%%%%%%
It perhaps bears repeating here that the $X_i$ are integration constants for the $\xi_i$ modes and break supersymmetry, while the $Y_i$ are integration constants for the modes $\phi^i$. One key result from this table which  cannot be gleaned from the field expansions we have provided, is that the mode $\xi_1$, whose integration constant is $X_1$ and which is the only mode  responsible for the force on a probe D3-brane, is the most convergent mode in the UV.

%%%%%%%%%%%%%%%%%%%%%%%%%%%%%%%%%%
\subsection{The Zero Energy Condition and Gauge Freedom} \label{sec:zeroenergy}
%%%%%%%%%%%%%%%%%%%%%%%%%%%%%%%%%%

In addition to the first-order equations of motion \eq{gradflow} there is a constraint, coming from the $R_{\tau\tau}$ component of Einstein's equation:
\be
\half G_{ab}\frac{d\phi^a}{d \tau} \frac{d\phi^b}{d \tau}= V(\phi)
\ee
which must be enforced in order for the solutions of our system of equations to correspond to solutions of supergravity. This constraint is imposed after fixing the gauge freedom associated with redefining the radial coordinate. When linearizing the PT ansatz around the KS background this becomes a constraint on just the $\xi_a$
\be
\xi_a \frac{d \phi_0^a}{d \tau}=0.
\ee
In addition,  the common rescaling of the coordinates $(t,x_1,x_2,x_3)$ can
be used to eliminate the integration constant in the metric function $A$ (since $A=\frac13 (-\tilde \phi_1 + \tilde \phi_4)$ in the UV this integration constant is  $Y_1^{UV}$). Hence the linearized solution depends on an even number of  physically-relevant parameters (fourteen), consistent with the expectation from the $AdS$-CFT correspondence.

On our solution to \eq{xieq} and \eq{phieq}, we find that the zero energy condition is a linear algebraic relation amongst the integration constants
\be
-6X_2 + 4 X_3+ X_4 + 3 X_5 -9 X_7=0 \label{zeroec}
\ee
(note that $X_1$ does not enter in the condition). Using this to eliminate $X_4$ or $X_5$ then leaves just two normalizable and two non-normalizable modes with $\Delta=4$ in Table 1.

%%%%%%%%%%%%%%%%%%%%%%%%%%%%%%%%%%
\section{Boundary Conditions for Three-Branes} \label{sec:bcs}
%%%%%%%%%%%%%%%%%%%%%%%%%%%%%%%%%%

Within the space of solutions we have derived in Section
\ref{sec:solution} we are interested in finding the modes which might
result from the backreaction of a collection of smeared anti-D3
branes. These branes are placed at $\tau=0$ and smeared on the finite
size $S^3$, and for describing them it is necessary to carefully
impose the correct infrared boundary conditions.

The gravity solution for a stack of localized D3-branes in flat space has a warp
factor $h(r)=1+Q/r^4$ and as $r\ra 0$ the full solution is smooth due
to the infinite throat. However when these branes are smeared in
$n$-dimensions, the warp factor will scale as $r^{-4+n}$ as $r\ra 0$
since it is now  the solution to a wave equation in dimension
$d=6-n$. This is the IR boundary condition we will impose on the
metric.

In addition we must impose boundary conditions on the various fluxes.
This is relatively straightforward for D3-branes in flat space, where
the energy from $F^{(5)}$ cancels against that from the curvature. In
the presence of other types of flux, the IR boundary conditions are
slightly more subtle. When the background is on-shell, contributions
to the stress tensor from all types of flux taken together cancel the
energy from the curvature: this is the basic nature of Einstein's
equation but this is too loose a criterion to signal the presence of
D3 branes.  The correct boundary conditions for D3-branes should require
the dominant contribution to the stress-energy tensor
to come from the $F^{(5)}$ flux.

%%%%%%%%%%%%%%%%%%%%%%%%%%%%%%%%%%
\subsection{BPS Three-Branes} \label{sec:BPS}

The Klebanov-Strassler background is a smooth solution and has no D3 Page charge. The D3 and D5 Page charges are
\bea
\cQ_{D3}^{Page}&=&\frac{1}{(4\pi^2)^2}\int_{\cM_5} \Blp \cF^{(5)}-B^{(2)} \w F^{(3)} +\half B^{(2)} \w B^{(2)}   \w F^{(1)} \Brp \non \\
\cQ_{D5}^{Page}&=&\frac{1}{4\pi^2}\int_{S^3} \Blp F^{(3)}-B^{(2)} \w F^{(1)} \Brp \non
\eea
and evaluated on the KS background these are
\be
\cQ_{D3}=0,\ \  \cQ_{D5}=P. \non
\ee

It is trivial to obtain the BPS solution in which smeared D3 branes
are added at the tip of the deformed conifold. In the PT ansatz one
can add D3-brane sources by shifting $\cF^{(5)}$ by a constant
multiple of the volume form on $T^{1,1}$. This shifts the D3 Page
charge, as well as $*\cF^{(5)}$, in such a way that the warp factor
becomes singular at the tip\footnote{Note that one can also shift the
  D3 Page charge by an integer multiple of $P$, by a large gauge
  transformation of $B_2$. However, this will not affect $*\cF^{(5)}$
  or the warp factor, and hence will leave the physics invariant. }.

The integral that gives the warp factor remains of the same form as in the original KS solution,
\be
h(\tau) \sim \int^\tau \frac{f_0(2P-F_0) + k_0F_0}{\sinh^2\tau' K^2(\tau')} d\tau'
\ee
with $K(\tau)=\frac{(\sinh 2\tau-2\tau)^{1/3}}{2^{1/3} \sinh \tau}$.
Shifting $\cF^{(5)}$ by a constant corresponds to shifting $f_0$ and
$k_0$ by equal amounts. Under a shift of $N/(2P)$ the D3 Page charge
shifts by
\beq
\cQ_{D3} \ra \cQ_{D3} + N
\eeq
while $\cQ_{D5}$ remains invariant. This introduces in the IR a
$1/\tau$ singularity in $h(\tau)$, whose physical interpretation is
obvious: we have smeared BPS D3-branes (whose harmonic function
diverges as $1/r^4$ near the sources) on the $S^3$ of the warped deformed conifold.

In order to ``calibrate'' our perturbation-theory machinery it is
interesting to see how this BPS solution can be obtained in an
expansion around the BPS KS background. First, all the $\xi_i$ fields,
that correspond to modes that break supersymmetry at first order,
should be set to zero, and hence $X_i=0$. Furthermore, since $\xi_i$ are zero,
$Y_2^{IR}=Y_2^{UV}$, and requiring no divergent terms in the IR of $\tphi_2$ sets these to zero.
The same thing happens with $Y_3$ and $Y_1$ (for the latter we require no constant term in the UV of
$\tilde \phi_1$). The constant $Y_8$ has a very clear meaning: it corresponds to a shift in the dilaton.
Note that this shifts $\tphi_5, \tphi_6$ and $\tphi_4$ as expected from the exact KS solution\footnote{The NSNS B-field
has a factor of $g_s$ which we have set for simplicity to $1$ in Eq. \eq{KSbackground}. Similarly the integral that gives $h$ in Eq. \eq{KSh} contains
a factor of $g_s^2$; shifting $g_s$ in the BPS solution changes $\tphi_4$ by a term proportional to $\tau/(4 \tau-1)$, exactly as in our UV expansions.}. We consider perturbation where the dilaton is not shifted at infinity and hence set
$Y_8^{IR}=Y_8^{UV}=0$. The fields $\tphi_5, \tphi_6$ and $\tphi_7$ then satisfy the homogeneous equation, and their
solution is obtained from (\ref{phi567}) by setting $\lambda_{5,6,7}$ to $Y_{5,6,7}$. Requiring no exponentially divergent terms in the UV
and no divergencies in the IR set respectively $Y_5=0$ and $Y_6=0$. The only non-zero
integration constant one can have is therefore $Y_7$.

Hence, the perturbation corresponding to adding $N$ BPS D3 branes is
obtained by just setting $Y_7 \sim N$. Note that this perturbation
causes the warp factor $\tphi_4$ to diverge in the infrared as $N/\tau$, exactly as
one expects from the full BPS solution. All the other $\tphi_i$ change
by subleading terms, except $\tphi_5$ and $\tphi_6$, which as argued above shift by
$N$.

Note that in the UV $Y_7$ multiplies a non-normalizable mode corresponding to the fact that we have changed the
rank of the gauge group of the dual field theory by adding D3-branes:
\be
\tphi^4 = - {2 Y_7 \over 3 P (4 \tau -1)} + ... \non
\ee
Hence the new warp factor is
\bea
h = e^{4A+4p-2x} &=& P^2 e^{-4 \tau /3 } (4 \tau -1) (1-2 \tphi^4) \non \\
&=&  P^2 e^{-4 \tau /3 } (4 \tau -1) +
{4 Y_7 P \over 3} e^{-4 \tau /3} + ... ~, \non
\eea
and we can see that $Y_7$ multiples a $1/r^4$ term, as one expects for pure (non-fractional) D3 branes.

%%%%%%%%%%%%%%%%%%%%%%%%%%%%%%%%%%
\subsection{Anti-BPS 3-branes in Anti-Klebanov-Strassler}

The anti-KS solution is obtained from the KS solution by flipping
the sign of $H^{(3)}$ and $\cF^{(5)}$ but not of $F^{(3)}$. This flips the signs of
$k_0$ and $f_0$ in the expression for  $H^{(3)}$ and has the same D5 brane
Page charge as the KS solution, namely $\cQ_{D5}^{Page}=P$. It is rather straightforward
to see that this solution is also supersymmetric but it preserves
different Killing spinors than the original KS solution. With the addition of $\ol{N}$ anti-D3 branes smeared on the $S^3$ the warp factor
$h$ again diverges in the infrared as
$\overline{N} \over \tau$, and the functions $f$ and $k$ are equal and approach  $-\overline{N} $. The D3 brane page charge is $\cQ_{D3}^{Page}=-\ol{N}$.

The KS solution with D3 branes at the bottom can be thought of as
coming from first taking the Ricci-flat deformed conifold with a
topological $F^{(3)}$ flux, puting D3 branes in, and backreacting this
configuration. The orientation of $\cF^{(5)}$ (and subsequently that of $H^{(3)}$) is determined
by the orientation of these D3 branes. In a similar fashion, one can obtain the anti-KS solution with anti-D3 branes
by starting from the same Ricci-flat deformed conifold with the same topological $F^{(3)}$ flux
and putting anti-D3 branes in; this backreacted system
has $\cF^{(5)}$ and $H^{(3)}$ of opposite orientation compared to the solution in section \ref{sec:BPS}.
Hence, for one orientation one obtains BPS D3 branes in the KS solution, while for
the opposite orientation one obtains the anti-BPS 3-branes in
anti-KS.

Note that if one thinks about the solutions in this way, it does not
make sense to a-priori fix the sign of the charge of the
``fractional'' D3 branes. The only quantity whose orientation is fixed
is the integral of $F^{(3)}$ on the 3-cycle. Hence, from this
perspective, changing the orientation of the D3 brane sources reverses
the sign of $\cF^{(5)}$ throughout the solution, and hence transforms
the ``fractional'' D3 branes into ``fractional'' anti-D3 branes.

%%%%%%%%%%%%%%%%%%%%%%%%%%%%%%%%%%
\subsection{Anti-BPS 3-branes in Klebanov-Strassler} \label{antiD3KS}

If one now inserts anti-D3 branes in the BPS Klebanov-Strassler
solution, one expects the physics at the tip to be dominated by the
anti-D3 branes sources, and in addition to have an extra D3-charge
dissolved in flux, as expected from the brane-flux annihilation
process \cite{Kachru:2002gs}.  As a result, we need to consider how
branes and fluxes contribute to the Maxwell charge. Recall that the
Maxwell charge for D3-branes is defined as \be
\cQ^{Max}_{D3}=\frac{1}{(4\pi^2)^2}\int_{T^{11}_\infty} \cF^{(5)},
\label{Qmax} \ee where we have indicated that the integral is over the
$T^{11}$ slice at infinity. For the zero-th order KS solution, the
Maxwell charge is not constant, it measures the running of the rank of
dual gauge groups and thus is a rough measure of the degrees of
freedom. However we are concerned only with the {\it additional}
Maxwell charge which we will add to the background and thus we will
abuse terminology and refer to this additional charge as
$\cQ_{D3}^{Max}$. In other words we have normalized the background
Maxwell D3 charge to zero.

As is clear from the Bianchi identity for $F^{(5)}$, $Q^{Max}_{D3}$ gets 
contributions from branes and flux
combined
\be
d\cF^{(5)}=H^{(3)}\w F^{(3)} + \sum_i \delta(x-x_i)
\ee
where the second term comes from explicit D3-brane sources.
Integrating \eq{Qmax} by parts yields
\bea
\cQ^{Max}_{D3} &=& q_b +q_f
\eea
where 
\bea
q_b&=&\frac{1}{(4\pi^2)^2}\int_{T^{11}_0} F^{(5)}, \label{qb}\\
 q_f&=& \frac{1}{(4\pi^2)^2}\int_{M_6} H^{(3)}\w F^{(3)}  \label{qf}
\eea
are the charges from the explicit D3-brane sources and flux
respectively. Note that $\cQ^{Max}_{D3}$ is gauge invariant (unlike
the Page charge) but is not necessarily quantized\footnote{A very nice
  description of the various charges in type II string theory can be
  found in \cite{Marolf:2000cb, Benini:2007gx, Aharony:2009fc}.}.

%%%%%%%%%%%%%%%%%%%%%%%%%%%%%%%%%%
\section{Constructing the Anti-D3 Brane Solution} \label{sec:antiD3}
%%%%%%%%%%%%%%%%%%%%%%%%%%%%%%%%%%

As advertized in Section \ref{sec:strategy}, we have two strategies to
look for the ``anti-D3 brane in KS'' solution. The first is to impose
UV boundary conditions, perturb the KS solution, and try to match the
expected boundary conditions in the infrared. The second is to impose
the IR boundary conditions (which make the near-tip solution to be
that of anti-BPS 3-branes in anti-Klebanov-Strassler), and to try to
obtain the UV KS boundary conditions as a non-normalizable
perturbation of the anti-KS solution.

%%%%%%%%%%%%%%%%%%%%%%%%%%%%%%%%%%
\subsection{UV $\ra$ IR, $X_1$ is non zero} \label{UVtoIR}
%%%%%%%%%%%%%%%%%%%%%%%%%%%%%%%%%%

The UV boundary conditions we wish to enforce on our prospective new
solutions are that $\cQ^{Max}_{D3}=\Nbar$ and that there should be a
non-zero force on a probe D3-brane.  In our ansatz the first boundary
condition requires a careful analysis of
\be
{\cal F}_5= (k F + f(2P -F)) \, \vol_{T^{11}}
\ee
in the UV. In addition the brane probe calculation of Section
\ref{forcesection} demonstrates that the force on a probe D3 brane
depends only on $X_1^{UV}$. Since $X_1^{IR}$ is proportional to
$X_1^{UV}$ (see Eq. (\ref{X1UVIR})), demanding a non-zero force in the
UV will imply
\beq \label{X1nonzero}
X_1^{UV} \neq 0 \ \Rightarrow X_1^{IR} \neq 0 \ .
\eeq
In addition, we also demand very divergent UV non-normalizable modes to be absent, and hence set 
\be \label{Y45UV}
 Y_5^{UV}= Y_4^{UV}= 0. 
\ee
These are the only UV conditions that we will impose so in principle all the 
other non-normalizable modes could be turned on.

Now we need to extract from this abstract discussion the explicit UV and IR behavior for the modes in our deformation space\footnote{Recall that in our notation $(\tphi_5,\tphi_6,\tphi_7)=(f_1,k_1,F_1)$, the perturbations to $(f_0,k_0,F_0)$.}.  The total Maxwell charge is evaluated in the UV by demanding
\be
(\tphi_6-\tphi_5)F_0+ (k_0-f_0)\tphi_7+2P\tphi_5= \cQ^{Max}_{D3} +\cO(e^{-\tau/3}).
\ee 
Using \eq{KSbackground}, \eq{Y45UV} and the UV expansions in section \ref{sec:phiUV} we find that
\bea
\cQ^{Max}_{D3} &=& P(\tphi^{UV}_5+\tphi^{UV}_6) \label{QD3UV} \non \\
&=& 2P \,Y_7^{UV}~. \label{QmaxUV}
\eea

In the IR, we want the $\tphi^i$ to contain no divergent terms except those coming from $q_b$.
From \eq{qb}  and \eq{PTfluxes} we immediately find that,
\be
(\tphi_6-\tphi_5)F_0+ (k_0-f_0)\tphi_7+2P\tphi_5 \sim q_b + \cO(\tau)~.
\ee
From the zero-th order expressions \eq{KSbackground} we see that
\be
F_0\sim \cO(\tau^2),\ \ f_0\sim \cO(\tau^3),\ \ k_0\sim \cO(\tau)
\ee
so that disallowing divergences in $(\tphi_5,\tphi_6,\tphi_7)$ we find the constraint
\be
2P\tphi_5 \sim q_b + \cO(\tau)~. \label{qbphi5}
\ee
Then on physical grounds we demand that in the IR the singularity in the Einstein tensor (due to the singularity in the warp factor) is commensurate with the singularity in the energy density from $F^{(5)}$ (recall that $\tphi_4$ is the warp factor):
\be
\tphi_4 \sim |q_b|/\tau + \cO(\tau^0). \label{qbphi4}
\ee
So this divergence is allowed since it is clearly coming from 
the anti-D3 brane sources.

However we must eliminate other divergences. We can see from the equations for $\tphi^2$ and $\tphi^3$ that this implies
\be \label{IRreg}
Y_2^{IR}= Y_3^{IR} = 0~.
\ee
Note that unlike the integration constant $X_1$, whose values in the
UV and IR expansions are proportional to each other, all the other 13
integration constants differ the UV and IR by an additive constant, as reviewed in section \ref{sec:X1eq0}. Hence, the fact that $Y_2^{IR}$
must be set to zero implies that $Y_2^{UV}$ will be a very nontrivial
function of all the $X_i$, which is highly unlikely to be zero unless
all the $X_i$ are zero and we have a BPS solution. Hence, any
prospective anti-D3 solution will source modes that behave in the UV
as $1/r^3$, contrary to what one might expect based on
codimension-counting.

The absence of divergences in $\tphi^7$ further implies that
\be
Y_6^{IR} = \frac{h_0}{3} (X_5^{IR} + X_{6}^{IR})
\ee
and the absence of $1/\tau$-divergences in $\tphi^6$ implies
\be
(X_5^{IR} + X_{6}^{IR}) = {16 (2/3)^{1/3} P X_1^{IR}\over h_0}.
\label{x5plus6}
\ee
Furthermore the absence of a log-divergent mode in $\tphi^6$ combined with the equations above also implies
\be
X_4^{IR}= -{ X_1^{IR}\over 6}.
\ee
The absence of a divergent term in $\tphi^8$ imposes a relation between $X_8$ and the other variables:
\be \label{IRreg2}
X_8^{IR} =-{P \over 6}(5X_5^{IR} - X_6^{IR} - 6 X_7^{IR}).
\ee
Furthermore, upon using (\ref{x5plus6}) one can see that the $1 \over \tau$ term in $\tphi^4$ depends only on $X_1^{IR}$ and $Y_{7}^{IR}$
\be
\tphi_4 = {1 \over \tau} \Blp -{2^{10/3} 3^{2/3} P\over h_0} Y_7^{IR} +  \frac{ 2^{4/3} }{3^{1/3}} X_1^{IR}  - {4096 P^4 \over
 3 h_0^2} X_1^{IR}\Brp + Y_4^{IR} +O(\tau) \label{tphi4IR}
\ee
Upon using these conditions above, the expansions of $\tphi_5$ becomes
\bea \label{phi56reg}
\tphi_5 &=& Y_7^{IR} + {256 P^3 \over 3 h_0} \Blp {2\over 3}\Brp^{2/3} X_1^{IR} +4(2/3)^{1/3} P X_1^{IR}\tau^2 ++\cO(\tau^3). \label{tphi5IR}
\eea
Note however that the values for $(Y_7^{IR}, X_1^{IR})$ have already
been fixed in terms of $\Nbar$: firstly, \eq{QmaxUV} fixes $Y^{UV}_7$
in terms of the number of initial anti-D3 branes ($\Nbar$) put into
the system, which is in turn related to $Y^{IR}_7$ by an additive
constant. Secondly, eqns.  \eq{qbphi4} and \eq{tphi4IR} give $X_1^{IR}$
in terms of $Y_7^{IR}$ and $q_b$.  One can then use \eq{qbphi5} and
\eq{tphi5IR} to obtain $X_1^{IR}$ in terms of $Y^{IR}_7$. At
this point we have computed all the IR quantities including $q_b$ in
terms of $\Nbar$ and one might declare this to be the solution for
anti-D3 branes in the KS background that we have been seeking.
However, the singularity analysis is more subtle than that presented
above.

\subsubsection{IR Singularities}
\label{Singularities}
 Recall that the NS and RR three forms in our ansatz are
\bea
 H_3&=&\tfrac12 (k-f) g_5 \wedge (g_1 \wedge g_3+ g_2 \wedge g_4) + d\tau \wedge (f' g_1 \wedge g_2+ k' g_3 \wedge g_4), \non  \\
  F_3&=&F g_1 \wedge g_2 \wedge g_5+ (2P-F) \, g_3\wedge g_4 \wedge g_5 +
 F' d\tau \wedge (g_1 \wedge g_3+ g_2 \wedge g_4) \ , \non
\eea
and in the IR the unperturbed metric is regular and is given by
\be
ds_{10}^2 \sim \alpha_1 ds_4^2 + \alpha_2\Blp\half d\tau^2 + (g_3^2+g_4^2+\half g_5^2)+ \frac{\tau^2}{4} (g_1^2+g_2^2)\Brp  \non .
\ee
for some numerical constants $(c_1,c_2)$. From this we observe that if
\be
(k-f)\sim a_0 + \cO(\tau) 
\ee
then the energy density has a term that divergences as
\be
H_3^2\sim \frac{a_0^2}{\tau^2}. \label{H2kf}
\ee
Furthermore, if 
\be
f\sim b_0 + b_1 \tau + b_2 \tau^2 + \cO(\tau^3) \label{fb0b1} 
\ee
then there is a singularity in the energy density of the form
\be
H_3^2 \sim \frac{b_1^2}{\tau^4} + \frac{b_2^2}{\tau^2}.  \label{H2f} 
\ee
The singularities from $a_0$ and $b_2$ are somewhat more benign than that from $b_1$, since their divergent energy density integrates to a finite action\footnote{We are very grateful to Igor Klebanov for sharing this observation with us.} (recall that  $\sqrt{G}\sim \tau^2+ \cO(\tau^4)$)
\be
|\int d^{10} x \sqrt{G} H_3^2 | <\infty.
\ee
The $b_1$ term in \eq{fb0b1} however would  result in a divergent action. Similarly, the IR expansion of $F_3$:
\be
\tphi_7= c_0 + c_1 \tau +\cO(\tau^2)
\ee
gives a singularity in the energy density of order
\be
F_3^2 \sim \frac{c_0^2}{\tau^4} + \frac{c_1^2}{\tau^2}  \label{F3sing} 
\ee
and so we see that both the RR and NS three-form field strengths have 
potentially inadmissible singularities.

Using the boundary conditions from section \ref{UVtoIR} we find that 
\bea
\tphi_5-\tphi_6 &=& -{16 P \over 3 } \Blp {2\over 3}\Brp^{1/3} X_1^{IR}+  \cO(\tau) \ , \\
\tphi_7&=& \frac{8P}{3} \Blp\frac{2}{3}\Brp^{1/3}  X_1^{IR} \, \tau + \cO(\tau^2).
\eea
The difference $\tphi_5-\tphi_6$ has a nonzero constant in its IR
expansion ($a_0\neq 0$), $\phi_5$ has a nonzero quadratic term
$(b_2\neq 0$) and $\tphi_7$ has a nonzero linear term ($c_1\neq 0$),
all of which are proportional to $X_1^{IR}$ and lead to an energy
density which diverges as $\tau^{-2}$. However, both the linear term
in $\tphi_5$ and the constant term in $\tphi_7$ are zero ($b_1=c_0=0$)
and thus there are no singular energy densities that scale like
$\tau^{-4}$, are hnce no infinite-action divergencies. 

Our analysis establishes that the only solution whose ultraviolet is
consistent with that of would-be anti-D3 branes in the
Klabanov-Strassler solution must have finite-action divergences in the
infrared, both in the RR and the NS-NS three-form field strengths. The
only way to eliminate these divergencies is by setting $X_1^{IR}=0$,
which in turn implies that $X_1^{UV}=0$ and is thus in contradiction
with the UV boundary condition \eq{X1nonzero} necessary in order to
have a nonzero force on a probe D3 brane.

Clearly, finite-action divergencies are better than infinite-action
divergencies, but it is not clear whether in the context of the
$AdS$-CFT duality even such finite-action divergencies should be
allowed. The best example of a finite-action singularity that must be
excluded from the spectrum of any consistent theory of two derivative
gravity is the negative-mass Schwarzchild solution\footnote{We are
  very grateful to Gary Horowitz for pointing this out.}.

Furthermore, this singularity does not appear a-priori to have any
direct connection with the infrared physics of probe anti-D3 branes in
the KS solution \cite{Kachru:2002gs}, and the brane-flux annihilation
that occurs via the polarization of these anti-D3 branes into NS5
branes wrapping an $S^2$ inside the $S^3$: First, brane-flux
annihilation results in the non-trivial division of $\cQ_{D3}^{Max}$
into $(q_b,q_f)$ and this in turn is related to the IR and UV values
of $f$, and not to the IR value of $k-f$. As such, this singularity
could not have been predicted as a consequence of brane-flux
annihilation. Second, the NS5 brane that mediates brane-flux
annihilation couples magnetically with an NS-NS three-form field
strength, while one of the fields whose energy density diverges in our
solution is a RR three-form field strength ($c_1\neq 0$). Third, in
the probe analysis of brane-charge annihilation, the amount of charge
dissolved in fluxes depends nonlinearly on the amount of anti-D3
branes, while in our solution both the anti-D3 charge and the
coefficient of the divergent terms are proportional to $X_1^{IR}$.

Hence, given the absence of a microscopic explanation for this
singularity, and given the absence of any criterion under which this
singularity would be acceptable while the negative-mass Schwarzschild
singularity would be not, it is fair to say that if this singularity
had appeared all by itself in a fully-backreacted solution, it should
have been deemed unacceptable.

However, as we have seen in the previous section, this singularity
does not appear all by itself: the infrared geometry also contains
metric and five-form fields whose energy density computed in
first-order perturbation theory diverges as $\tau^{-4}$. These fields
have an obvious microscopic interpretation: they are sourced by a
smeared distribution of anti-D3 branes, the metric and five-form
divergences are commensurate, as typical for D-branes, and we expect
this singularity to make perfect sense in string theory. Hence, it is possible 
that upon taking into account the string-theoretic
resolution of the singularity sourced by the smeared D3-branes, the
the weaker finite-action singularity may become benign.

%%%%%%%%%%%%%%%%%%%%%%%%%%%%%%%%%%
\subsection{IR $\ra$ UV}

We now start from a solution that eliminates the IR singularities of
the previous section, and argue that the UV expansion of this solution
cannot match the UV expansion of the KS solution with none but the
charge non-normalizable mode turned on. This problem is equivalent to
starting from the BPS D3 branes in the IR and arguing that one cannot
match the anti-KS solution in the UV.

If we are to match the boundary conditions near the anti-branes, Eqs. (\ref{IRreg})-\eq{IRreg2},
the fields $\tphi^5$ and $\tphi^6$ must be equal, and using (\ref{phi56reg}) this implies
\be
 X_1^{IR} = X_1^{UV} = 0. \non
\ee
Note that in the absence of $X_1$ we have already solved exactly the
equations (\ref{xieq}) for the $\xi^i$ in Section \ref{solvingxi}, and
thus the values of the $X_i$ in the UV and in the IR must be the same.
The infrared regularity equations in the previous section imply:
\bea
X_5 + X_{6}= X_4 &=& 0~, \non \\
Y_2^{IR}= Y_3^{IR}=Y_6^{IR}&=&0~, \label{Y23IR}
\eea
and therefore
\be
X_8= -{P}(X_5-X_7)~. \non
\ee

Furthermore, since we do not want divergent and non-normalizable $1/r$ modes in the UV,
\be
X_3=X_2=X_7=0~. \non
\ee
However, in deriving the infrared boundary conditions in the previous
subsection we have used the zero energy condition (\ref{zeroenergy})
to express $X_2$ as a function of $X_3,X_4,X_5$ and $X_7$, and setting
it to zero combined with the other equations implies
\be
X_5 = 0~ \non
\ee
and therefore all $X$'s must vanish.
Hence, if one tries to match all the correct IR boundary conditions
and avoid un-physical non-normalizable modes, one must set all the
$\xi_i$ to zero.

One can now proceed exactly as in Section \ref{solvingphi} to
establish that $\tphi_8$ is a constant, and use equations \eq{solphi2}, \eq{solphi3} and
 \eq{solphi1} together with \eq{Y23IR}
to show that $\tphi^2=\tphi^3=0$, and that $\tphi^1$ must be a constant.
The next step is to solve the system \eq{phi567} exactly, and
match the infrared and ultraviolet integration constants. One finds
that
\bea
\tphi^5\!&=&\!\frac{1}{2} {\rm sech}^2\frac{\tau}{2} \Blp\cosh\tau (2 Y_5 \tau- Y_5 \sinh\tau+Y_7 -P Y_8 \tau
  )\non \\
  && \hskip 3cm +\sinh \tau (P Y_8-2Y_5)+ Y_5 \tau + Y_6 + Y_7\Brp, \nonumber \\
\tphi^6\!&=&\! -\frac{1}{2} \csch^2\frac{\tau}{2} \Blp
- \cosh\tau (2 Y_5 \tau+ Y_5 \sinh\tau+Y_7 -P Y_8 \tau
  ) \\
&~& \hskip 3cm+~\sinh \tau (P Y_8-2Y_5)+ Y_5 \tau + Y_6 + Y_7\Brp, \nonumber\\
\tphi^7\!&=&\!-Y_5 \cosh \tau + (Y_5 \tau - Y_6 )\csch \tau \nn \ .
\eea
Comparing to the expansions in Sections \ref{sec:phiIR} and \ref{sec:phiUV} we get
\be
Y_5= Y_5^{UV}=Y_5^{IR}+ {P Y_8 \over 6}~, ~~Y_6=Y_6^{IR}= Y_6^{UV}~, ~~Y_7=Y_7^{IR}= Y_7^{UV}~.
\ee
Requiring no divergencies in the UV and IR sets
\be
Y_5= Y_6=0 ,
\ee
Importantly, we see that the sign of $Y_7$, which is proportional to
the effective D3-brane charge (recall that $Y_8$ just corresponds to a
shift in the dilaton), cannot change from the IR to the UV. This
implies that this solution has positive D3 brane charge throughout
Hence, the solution that preserves D3 boundary conditions in the
infrared cannot match a solution with anti-KS boundary conditions in
the UV and viceversa.

\vskip 1cm

  %%%%%%%%%%%%%%%%%%%%%%%%%%%%%%%%%%
\section{Discussion} \label{disc}
  %%%%%%%%%%%%%%%%%%%%%%%%%%%%%%%%%%

With a view towards identifying the modes sourced by anti-D3 branes,
we have computed the spectrum of linearized fluctuations around the
Klebanov-Strassler solution which respect the symmetries of the
original background. Finding such solutions, with anti-D3 brane
boundary conditions in the infrared and Klebanov-Strassler boundary
conditions in the ultraviolet has been a key unresolved step in many
constructions of de Sitter vacua in string theory. Much work has been
done using this mechanism to show that large numbers of such vacua
exist.

We have found in our spectrum a single solution where some of the
anti-D3 charge is dissolved in flux, supersymmetry is broken (as
suggested in \cite{Kachru:2002gs}) and there is a force on a probe D3
brane of the same form as the one computed in \cite{Kachru:2003sx}.
While this is all somewhat encouraging, the solution we find must
necessarily have an infrared singularity, coming from the fact that the energy
densities of both the RR and NS three form fluxes diverge
quadratically at the origin. 

We have discussed this singularity in great detail in Section
\ref{UVtoIR}, and have seen that as far as first-order
perturbation theory is concerned, this singularity is better behaved
than other possible IR singularities - in particular its divergent
energy density integrates to a finite action. However, even if the finiteness of the action 
makes this singularity more acceptable than others, it cannot be used as a criterion to establish that this
singularity is physical. For example, the negative-mass Schwarzschild solution also has a finite-action singularity, and yet this solution is unphysical, and must be eliminated from the bulk spectrum in order for the gauge/gravity duality to make sense. Indeed, the breakdown of general
relativity at a singularity does not come from the diverging action
but rather from the fact that the derivative expansion in the
effective action for gravity does not provide a good perturbation
theory. Hence, divergent energy densities in the RR and NS-NS fields
are as unphysical as a divergent Ricci scalar, and indicate that
perturbation theory has broken down.

We have also tried to find a microscopic interpretation for this 
singularity, to see whether the divergent fields can be explained as
coming from D-brane or NS5-brane sources that one may expect to find
in the infrared, and hence whether stringy corrections could resolve
this singularity.  We have argued that the existence of this
singularity is neither a direct consequence of the expected brane-flux
annihilation, nor of the polarization of anti-D3 branes into NS5
branes in the infrared.  From such arguments, the most direct
conclusion would be that this finite-energy singularity should be treated as other
such singularities in gravity and must be removed.

On the other hand there is one reason that we should persist with this
solution and perform further tests on it despite the singularity. It
may be that the finite-action singularity is not a {\it signal} of the breakdown of
general relativity but rather a {\it consequence}.  It is clear that
one can never completely capture the physics of the solution near the
anti-D3 branes by doing perturbation theory around the original KS
background, since in that region the energy in the fields sourced by
the explicit anti-D3 brane sources diverges. This is similar to the
fact that one cannot treat an electron in a background electric field
as a perturbation, since near the electron its electric field always
dominates. The energy density of the fields sourced by the explicit anti-D3 brane
sources in the IR diverges quartically in the IR, whereas the energy density of the three
forms diverges quadratically. It may be that the stringy resolution of
the anti-D3 brane singularities, or even their full supergravity backreaction, will also resolve the milder finite-action 
singularity we have found and render the whole solution physical. 

It is clearly crucial to establish whether this finite-action singularity is indeed
resolved this way and could be acceptable from the point of view of string
theory, or whether it is not acceptable and must be removed. 

If the finite-action singularity is physical, than our analysis has found a
first-order backreacted solution that has anti-D3
brane charge at the bottom of the Klebanov-Strassler solution. The objects with 
anti-D3 brane charge source modes that in the ultraviolet can be treated as a perturbation
of the Klebanov-Strassler solution, and it is very likely that this solution represents the
backreaction of the smeared system of anti-D3 branes considered by
Kachru Pearson and Verlinde in \cite{Kachru:2002gs}. Our analysis does not fully
establish that these anti-D3 branes source only normalizable
modes - to prove this one needs to relate the UV and IR integration
  constants, which is quite nontrivial \cite{BGHwip}. If
this is so, then this solution would represent the supergravity dual
of the conjectured metastable vacuum of the Klebanov-Strassler field
theory, and would be the first supergravity solution dual to a
metastable vacuum of a four-dimensional theory. By exploring the vacuum expectation values
of various operators, this solution would allow one to characterize
very precisely the physics of this metastable vacuum, and would quite likely increase our understanding of gauge theory metastable vacua in general.

On the other hand, if this singularity is not physical, our analysis
establishes that the solution corresponding to backreacting anti-D3
branes at the bottom of the Klebanov-Strassler background cannot be
treated in the UV as a perturbation around the BPS vacuum. Hence,
these anti-D3 branes source non-normalizable modes (other than the charge) whose energy is at
least as strong as that of the original KS solution.

While this conclusion might seem at first glance unexpected, it is is fact not once one realizes that some of the modes
may have two-dimensional dynamics, and a perturbation is such modes does not decay at infinity. Indeed, as we have argued in the Introduction, all supergravity solutions dual to four-dimensional theories with a running coupling constant must have bulk modes that depend logarithmically on the radial direction, coming from the fact that the coupling constants of these theories run logarithmically with the energy. Hence, it is quite natural to expect the anti-D3 branes to couple to those modes, and hence to source perturbations that grow logarithmically with the radius. It is not hard to see that the energy of such logarithmic modes is as strong as that of the unperturbed solution. 

Furthermore, this behavior would mimic exactly what happens when one tries constructing an MQCD metastable vacuum
\cite{Bena:2006rg}. First, if one tries to obtain this vacuum by perturbation theory around the BPS one, one also finds modes that diverge in the infrared, and cannot match the correct boundary conditions. This is a first
hint that the fully-backreacted metastable vacuum might not exist, and is confirmed by the 
subsequent backreacted calculation. Furthermore, if one tries to find the solution whose infrared has regular boundary conditions, one finds a solution that differs from the BPS one by a logarithmically-growing mode, and that cannot be obtained as a perturbation of the latter. This is again a reflection of the fact that this mode has two-dimensional dynamics, and once excited in the infrared it does not decay at infinity. 

If one takes the analogy with the MQCD construction of metastable vacua one step further, one can also identify an obvious candidate for the full solution consistent with anti-D3 branes in the infrared and no finite-action singularity: 
the anti-D3 branes in the anti-Klebanov-Strassler solution. Much like the MQCD IR-compatible solution\footnote{Given by Eq. (3.22) in
  \cite{Bena:2006rg}.}, this solution differs from the BPS one by
log-growing modes, it is not a perturbation of the latter and is
BPS by itself.  If this is indeed {\it the} solution for the
back-reacted anti-D3 branes, this would imply that the sign of the
charge dissolved in flux in the KS background must always have the
same sign of the D3 branes at the bottom, and changes sign as one changes the orientation of these branes. 

If the finite-action singularity is excluded, or if it is allowed 
but the anti-D3 branes source modes that are nonnormalizable, this would strongly suggest
that the dual gauge theory does not have a metastable vacuum. Though unexpected, this result would be 
consistent with the fact that a careful analysis of the vacuum structure of this theory 
\cite{Dymarsky:2005xt} has not found such a vacuum. Furthermore, as we have explained in the Introduction, 
this would probably affect the KKLT construction of deSitter vacua in string theory, and the existence 
of a landscape of such vacua in general, which in turn may invalidate many string
phenomenology and string cosmology constructions.

It is fair to say that neither our analysis, nor any in the literature definitively show that the finite-action singularity is unphysical, and hence rule out the existence of KS anti-D3 metastable vacua. On the other hand, we have not been able to find  any microscopic arguments or singularity-resolution arguments of why this singular behavior in the fields may be acceptable, which would establish that the solution we have constructed is the supergravity dual of the KS metastable vacuum, and implicitly prove that such a vacuum exists.

Our analysis crystalizes and reduces the complicated problem of the existence of KS metastable vacua to the simpler problem of whether the finite-action singularity is physically acceptable or not.  Establishing whether this singularity is physical would ideally come from a microscopic understanding of the infrared physics or from a near-brane analysis of the fully backreacted, localized solution (for example extending the analysis of \cite{DeWolfe:2004qx}). Any developments on these issues would clearly be very exciting.

\vskip 1cm

  %%%%%%%%%%%%%%%%%%%%%%%%%%%%%%%%%%
\noindent {\bf Acknowledgements}:
\noindent We would like thank O. Aharony, C. Charmousis, A. Dymarsky, M.
Gomez-Reino, S. Kachru, I. Klebanov, G. Horowitz, S. Kuperstein, J. Maldacena, J. Polchinski, N. Seiberg, D. Shih, F. Vernizzi and E. Witten for useful discussions and particularly H. Verlinde for challenging debates regarding the infra-red physics. N.H. would like to thank Princeton University for hospitality. The work of IB, MG and NH was supported in part by the DSM CEA-Saclay, by the ANR grants BLAN 06-3-137168 and JCJC ERCS07-12 and by the Marie Curie IRG 046430.
 %%%%%%%%%%%%%%%%%%%%%%%%%%%%%%%%%%

%%%%%%%%%%%%%%%%%%%%%%%%%%%%%%%%%%
 %%%%%%%%%%%%%%%%%%%%%%%%%%%%%%%%%%
 \begin{appendix}
 %%%%%%%%%%%%%%%%%%%%%%%%%%%%%%%%%%
%%%%%%%%%%%%%%%%%%%%%%%%%%%%%%%%%%
\section{Conventions}  \label{conventions}
%%%%%%%%%%%%%%%%%%%%%%%%%%%%%%%%%%
We use the somewhat unsightly notation $\xi_{a\cite{Kuperstein:2003yt}}$ to indicate that we refer to $\xi_a$ as defined in reference \cite{Kuperstein:2003yt}. This is to try to keep straight the different conventions which appear in the various papers on the subject . The relation to the variables used in other papers is the following. The angular variables  $\epsilon_1,\epsilon_2,\epsilon_3,e_1,e_2$ of \cite{Papadopoulos:2000gj} are related to $g_i$ by
\beq
\epsilon_1= \tfrac{1}{\sqrt2}(-g_2+g_4) \, , \  \epsilon_2= \tfrac{1}{\sqrt2}(-g_1+g_3) \, , \  e_1= \tfrac{1}{\sqrt2}(g_2+g_4) \, , \ e_2= \tfrac{1}{\sqrt2}(g_1+g_3) \, , \ \epsilon_3=g_5 \ .
\eeq
Our radial variable $\tau$ is related to the radial variables in \cite{Klebanov:2000hb,Papadopoulos:2000gj,Kuperstein:2003yt,Borokhov:2002fm,DeWolfe:2008zy}
 by
 \beq
 d\tau=d\tau_{\cite{Kuperstein:2003yt}}=e^{4 p_1 } d\tau_{\cite{Borokhov:2002fm}}=e^{4 p_1 }d\tau_{\cite{Klebanov:2000hb}}=e^{4(p_0+p_1)} du_{\cite{Papadopoulos:2000gj}}=\sqrt6 e^{3p_0+x_0+4p_1} d\,  \rm{ln} r_{\cite{DeWolfe:2008zy}} \ .
 \eeq
To zeroth order in perturbation this last relation reads in the UV $r=e^{\tau/3}$.

Our metric functions $x, y, p, A$  are related to  $q, y, p, Y$ used in \cite{Kuperstein:2003yt} and $a, b$ ($y=0$ and in the solution $c=0$) in  \cite{DeWolfe:2008zy}  (relations to the latter are only valid in the UV) by
 \bea
&& x=3q+2 p_{\cite{Kuperstein:2003yt}} = -2a-\rm{ln} 6\, , \ y=y_{\cite{Kuperstein:2003yt}}\ ,\non \\
&& \ p= p_{\cite{Kuperstein:2003yt}}-q=\frac13(2a-b+\rm{ln} (3\sqrt6)) \, , \ A=Y=a+\rm{ln}r \ .
 \eea

The functions $f,k,F$ in the 3-form fluxes are related by
\bea
&&f=-2P f_{\cite{Kuperstein:2003yt}}=2 f_{\cite{Borokhov:2002fm}}=6 k_{\cite{DeWolfe:2008zy}} \ ,\non \\
&& k=-2P k_{\cite{Kuperstein:2003yt}}=2 k_{\cite{Borokhov:2002fm}}=6 k_{\cite{DeWolfe:2008zy}} \ , \\
&& \ F=2P F_{\cite{Kuperstein:2003yt}}=2 F_{\cite{Borokhov:2002fm}}.  \non
\eea

%%%%%%%%%%%%%%%%%%%%%%%%%%%%%%%%%%%%
\section{Behavior of $\txi$} \label{xiasympt}
%%%%%%%%%%%%%%%%%%%%%%%%%%%%%%%%%%%%

%%%%%%%%%%%%%%%%%%%%%%%%%%%%%%%%%%%%
\subsection{IR behavior of $\txi$} \label{app:IRxi}
%%%%%%%%%%%%%%%%%%%%%%%%%%%%%%%%%%%%

The IR behavior of $\txi_a$ is the following
\bea
\txi_1 & =& X_1^{IR}  \left(1-\frac{16 \left(\frac{2}{3}\right)^{1/3} P^2 \, \tau^2}{3 \, h_0}\right)  +{\cal O}(\tau^4) \nn \\
\txi_4 & =& X_4^{IR} +\frac{16 \, X_1^{IR} \left(\frac{2}{3}\right)^{1/3} P^2 \, \tau^2}{3 \, h_0}  +{\cal O}(\tau^4) \nn \\
\txi_5&=& \frac{8\, X_1^{IR} \left(\frac{2}{3}\right)^{1/3} P }{  h_0 \, \tau}+X_5^{IR} +  \frac{16 \, X_1^{IR} \left(\frac{2}{3}\right)^{1/3} P^2 \, \tau}{15 \, h_0} +{\cal O}(\tau^4) \nn \\
\txi_6 &=& \frac{X_5^{IR}+X_6^{IR}}{2 \tau} + X_5^{IR} + \left( \frac{56 \,  X_1^{IR} \left(\frac{2}{3}\right)^{1/3} P^2 }{15 \, h_0} -\frac{X_5^{IR}+X_6^{IR}}{12}\right) \tau
+{\cal O}(\tau^2) \\
\txi_7 &=& \left(- \frac{8  X_1^{IR} \left(\frac{2}{3}\right)^{1/3} P^2 }{  h_0} +\frac{X_5^{IR}+X_6^{IR}}{2}\right) \frac{1}{\tau^2}\non \\
&& +
\left( -  \frac{4 X_1^{IR} \left(\frac{2}{3}\right)^{1/3} P^2 }{3 \, h_0} +\frac{X_5^{IR}+X_6^{IR}}{12}\right) +\frac{X_5^{IR}-4 X_7^{IR}}{3} \tau + {\cal O}(\tau^2) \nn \\
\txi_8 &=& X_8^{IR} + \frac16 P (5 X_5^{IR} - X_6^{IR} -6 X_7^{IR}) \non \\
&&+ P \left(   \frac{32 X_1^{IR} \left(\frac{2}{3}\right)^{1/3} P^2 }{45 \, h_0} +\frac{X_5^{IR}+X_6^{IR}}{15}\right) \tau^2 + \frac{2}{9} P (X_5^{IR}-X_7^{IR}) \tau^3 +  {\cal O}(\tau^4) \nn \\
\txi_3 &=& -\frac53 X_1^{IR} + \frac{16 X_1^{IR} \left(\frac{2}{3}\right)^{1/3} P^2 \, \tau^2}{3 \, h_0} + \frac{8}{3} X_3^{IR} \tau^3 + {\cal O}(\tau^4) \nn \\
\txi_2 &=& \left( \frac{ X_1^{IR}}{9} +   \frac{16  X_1^{IR} \left(\frac{2}{3}\right)^{1/3} P^2 }{3 \, h_0} +\frac{X_4^{IR}- P(X_5^{IR}+X_6^{IR})}{3}\right)  \non \\
&&+\left(-\frac13 X_4^{IR}+2X_2^{IR}-P X_5^{IR} - \frac43 X_3^{IR}+ 3 P X_7^{IR} \right) \tau \non \\
&&+  \left( \frac{X_1^{IR}}{18}  +   \frac{56 X_1^{IR} \left(\frac{2}{3}\right)^{1/3} P^2 }{15 \, h_0} +\frac{X_4^{IR}+ \frac35 P(X_5^{IR}+X_6^{IR})}{6}\right) \tau^2 +  {\cal O}(\tau^3) \nn
\eea

%%%%%%%%%%%%%%%%%%%%%%%%%%%%%%%%%%%%
\subsection{UV behavior of $\txi$} \label{App:UVxi}
%%%%%%%%%%%%%%%%%%%%%%%%%%%%%%%%%%%%

The UV behavior of $\tilde \xi_a$ is the following

\bea
\tilde \xi_1&=& X_1^{UV} e^{-4 \tau /3} (3 - 12 \tau) + {\cal O}(e^{-10 \tau/3}) \nn \\
\tilde \xi_4&=&  X_4^{UV} - X_1^{UV} \left(3 - 12 \tau \right) e^{-4 \tau /3} + {\cal O}(e^{-10 \tau/3}) \nn \\
\tilde \xi_5&=&  X_5^{UV} -\frac{2}{P} X_1^{UV} e^{-4 \tau/3} + {\cal O}(e^{-10 \tau/3}) \nn \\
\tilde \xi_6&=& \frac12 X^{UV}_7 e^\tau + \left(2 \tau( X^{UV}_5-X^{UV}_7) +  X_5^{UV} + X_6^{UV} + \tfrac{1}{2} X_7^{UV}\right) e^{-\tau} + {\cal O}(e^{-3\tau})\nn \\
\tilde \xi_7 &=& - \frac12 X_7^{UV} e^\tau + \left(2 \tau(  X_5^{UV}-X_7^{UV}) - X_5^{UV} + X_6^{UV} + \tfrac{5}{2} X_7^{UV} \right) e^{-\tau} + {\cal O}(e^{-3\tau})  \\
\tilde \xi_8 &=& X_8^{UV} + P  (X^{UV}_5-X^{UV}_7) \tau+ \frac32 X_1^{UV} e^{-4 \tau/3}  \nn \\
&& +2 P \Blp -2  (X_5^{UV}-X_7^{UV})\tau^2 +  (3  X_5^{UV} -X_6^{UV} -4 X_7^{UV})\tau+X_6 +X_7^{UV} \Brp e^{-2\tau} +  {\cal O}(e^{-3\tau}) \nn \\
\tilde \xi_3&=& X_3^{UV} \left( e^{2\tau} - 4 \tau -e^{-2\tau} \right) + \frac{3}{25} X_1^{UV} e^{-4 \tau/3} (-23 + 140 \tau) + {\cal O}(e^{-10 \tau/3}) \nn \\
\tilde \xi_2&=& \left((-\frac23 X^{UV}_3+P X^{UV}_7)\tau  +  X^{UV}_2 \right) e^{\tau} \nn \\
&& \!\!\!\!\!\!\!\!\!\!\!\!\!\!\!\!\!\! + \left(   (-\frac23  X^{UV}_3\! +\! 5 P X^{UV}_7\! -\! 4P X^{UV}_5) \tau \! -\! X^{UV}_2 \!+\! \frac13 X^{UV}_4 \!+\! P (X^{UV}_5\! -\! X^{UV}_6\! -\! 2 X^{UV}_7) \right) e^{-\tau} \! +\! {\cal O}(e^{-3\tau}) \nn
\eea

\end{appendix}
%%%%%%%%%%%%%%%%%%%%%%%%%%%%%%%%%%%%
%%%%%%%%%%%%%%%%%%%%%%%%%%%%%%%%%%%%
%\bibliographystyle{utphys} 
\bibliography{AntiD3revision}

%%%%%%%%%%%%%%%%%%%%%%%%%%%%%%%%%%%%
\end{document}